\documentclass[%
 reprint,
superscriptaddress,
 amsmath,amssymb,
 aps,
]{revtex4-2}

\usepackage{graphicx}% Include figure files
\usepackage{dcolumn}% Align table columns on decimal point
\usepackage{bm}% bold math
\usepackage{enumitem}
\begin{document}

\preprint{APS/123-QED}

\title{Superradiance from a Relativistic Source}% Force line breaks with \\

\author{C. M. Wyenberg}
 \email{cwyenber@uwo.ca}
 \affiliation{Department of Physics and Astronomy, The University of Western Ontario, London, Ontario, Canada, N6A 3K7}
%  \altaffiliation[Also at ]{Physics Department, XYZ University.}%Lines break automatically or can be forced with \\

\author{F. Rajabi}
 \affiliation{Perimeter Institute for Theoretical Physics, Waterloo, Ontario, Canada, N2L 2Y5}
 \affiliation{Institute for Quantum Computing and Department of Physics and Astronomy, The University of Waterloo, Waterloo, Ontario, Canada, N2L 3G1}
 \affiliation{Department of Physics and Astronomy, The University of Western Ontario, London, Ontario, Canada, N6A 3K7}
 
\author{M. Houde}%
 \affiliation{Department of Physics and Astronomy, The University of Western Ontario, London, Ontario, Canada, N6A 3K7}

% \collaboration{MUSO Collaboration}%\noaffiliation

%  \homepage{http://www.Second.institution.edu/~Charlie.Author}
% \affiliation{
%  Second institution and/or address\\
%  This line break forced% with \\
% }%
% \affiliation{
%  Third institution, the second for Charlie Author
% \affiliation{%
%  Authors' institution and/or address\\
%  This line break forced with \textbackslash\textbackslash
% }%

% \collaboration{CLEO Collaboration}%\noaffiliation

\date{\today}% It is always \today, today,

\begin{abstract}

We construct a model of cooperative superradiant emission from a highly relativistic multi-particle source. We revise an existing model of the literature for a relativistic two-level particle, and construct from it a Hamiltonian describing relativistic velocity dependent multi-particle superradiance. We adapt the standard diagrammatic framework to compute time evolution and density operators from our Hamiltonian, and demonstrate during the process a departure from standard results and calculation methods. In particular, we demonstrate that the so-called vertical photon result of the literature is modified by the relativistic Lorentz factor of the sample; we also introduce a set of coupled differential equations describing certain propagators in the velocity-dependent small sample framework, which we solve numerically via a hybrid fourth order Runge-Kutta and convolution approach. We demonstrate our methods for the simple case of two highly relativistic particles travelling with slightly differing velocities simulated at varying relativistic mean sample $\beta$ factors, and evaluate velocity coherence requirements for a sample to demonstrate enhanced superradiant emission in the observer frame. We find these coherence requirements to become increasingly restrictive at higher $\beta$ factors, even in the context of standard results of relativistic velocity differential transformations.

% \begin{description}
% Secondary publications and information retrieval purposes.
% You may use the \texttt{description} environment to structure your abstract;
% use the optional argument of the \verb+\item+ command to give the category of each item. 
% \end{description}
\end{abstract}

\maketitle

%\tableofcontents

\section{\label{sec:intro}Introduction}

In the quantum optics phenomenon of superradiance (SR) a collection of excited particles evolves into highly entangled states which generate greatly enhanced radiative emission over that predicted by the spontaneous decay of its constituent particles. In his seminal theoretical treatment of SR, Robert Dicke \citep{Dicke1954} first described the enhanced emission process as a cascade through symmetric states isomorphic to the maximal total angular momentum states from the elementary theory of the addition of spin-$1/2$ particles. If $|S\left(N,m\right)\rangle$ denotes the symmetric superposition of the possible quantum states of $N$ particles having $m$ excited and $N-m$ in the ground state (for example, $|S\left(2,1\right)\rangle = \left(|e,g\rangle + |g,e\rangle\right)/\sqrt{2}$), then the Dicke model describes the SR emission process by the cascade
\begin{equation}
    |S\left(N,N\right)\rangle \rightarrow |S\left(N,N-1\right)\rangle \rightarrow \dots \rightarrow |S\left(N,0\right)\rangle.
\end{equation}

The probability of transition $\mathcal{P}_{k,k-1}=|S\left(N,k\right)\rangle \rightarrow |S\left(N,k-1\right)\rangle$ in the Dicke model is computed from a perturbation analysis of the Hamiltonian coupling the particle to the quantized radiation field via the relevant transition operator (electric dipole, magnetic dipole, or other), from which the central state $k=N/2$ \footnote{$k=\left(N\pm1\right)/2$ for $N$ odd} is found to couple most strongly to the radiation field. The full cascade through all states is characterized by the two salient SR features of a delay and enhancement in peak emission intensity. These theoretical predictions have been well verified by laboratory demonstrations of SR \citep{Skribanowitz1973} since their original prediction by Dicke. Additionally, more sophisticated theoretical models of SR have been developed which generalize the Dicke model to extended volumes; which introduce realistic dephasing, relaxation, and pumping terms; and which lend themselves to more efficient numerical simulation. SR indeed possesses a rich experimental and theoretical history at the scale of laboratory quantum optics.

SR has also been recently applied to astrophysical environments, where it has been used to model maser flares \citep{Rajabi2016a, Rajabi2016b, Rajabi2019, Rajabi2020a} and fast radio bursts \citep{Rajabi2017, Houde2019}; however, such environments lead to regimes of SR not realized in the laboratory and thus as of yet untreated theoretically. Some initial progress has been very recently made towards the theory of wide inhomogeneous broadening of SR, and two $\mathcal{O}\left(n\right)$ complex algorithms for simulating transients of $n$ velocity channels have been derived \citep{Wyenberg2021, Wyenberg2022} and applied \citep{Wyenberg2022} to astrophysical environments; however, there presently exists in the literature no relativistically covariant formulation of SR. Such a model could prove essential to analyzing, for example, the proposed highly relativistic SR events of \cite{Rajabi2020b}. The objective of this work is to develop a simple model of SR emission from a small sample of particles travelling with differing relativistic velocities, and with it to determine the degree of velocity coherence required in order to measure enhanced SR emission in the observer's frame.

There exist various representations of SR which suggest differing approaches to this problem. In the so-called Maxwell-Bloch model of a large population in an extended medium, for example, the complicated quantum mechanical time evolution is shown \citep{Gross1982, Benedict1996, MacGillivray1976} to very quickly transition into a stage described by an ensemble of population inversion, polarization, and electric field trajectories determined by a relatively simple set of partial differential equations. In this later stage a correspondence to the classical electric and magnetic fields may be made, which could potentially be transformed to the observer's frame by the usual relativistic transformation of the electromagnetic field tensor \citep{Rajabi2020b}.

As a simple and fundamental first step towards a relativistic description of SR, however, this work does not seek to model a large population. Moreover, it is not a simple exercise to generalize the aforementioned result \citep{Gross1982, Benedict1996, MacGillivray1976} regarding the emergence of an ensemble of trajectories to observers of a relativistic source. We seek rather to develop a relativistic model of the proper quantum mechanical evolution of a small number of particles capable of SR emission.

Constructing a simple relativistic model of the quantum evolution of even a small number of particles presents numerous challenges. First, observer frame measurements cannot be described by the classical relativistic transformation of the electromagnetic field tensor. Instead, the time evolution of the observer frame quantum state must be computed from a Hamiltonian built upon a covariant formulation of the matter-field interaction. For this purpose we first introduce in Section \ref{sec:covariant} an existing covariant model \citep{Boussiakou2002} of a two-level particle established in the literature.

The second challenge faced by a relativistic model is the interpretation of the cascade through symmetric states. As detailed momentarily in Section \ref{sec:existing}, the small sample Dicke model \citep{Dicke1954} identifies the intensity transient with the average emission rate over a distribution of stochastic cascades, where each realization involves transitions between the symmetric states at definite times \footnote{This method is, strictly speaking, not theoretically rigorous; nonetheless, the more sophisticated diagrammatic methods of quantum optics confirm its results.}. In the case of photons emitted from a relativistic source, however, Dicke's interpretation becomes nuanced when considerations of time dilation or other relativistic effects are introduced into the stochastic cascade analysis. We therefore use the full diagrammatic method of quantum optics to evolve the matter-field system from the covariantly-derived Hamiltonian, and depart from the concept of stochastic discrete-time transition events in favour of a purely mathematical description of the system's time evolution operator.

The third challenge involves the choice of basis for the Hilbert space describing each particle's center of mass coordinate. SR is highly dependent upon the relative position of the particles, which therefore suggests the use of a position basis; however, we are interested in velocity coherence effects which are more naturally realized in a momentum basis. We define a compromised representation of the problem in Section \ref{subsec:compromised_basis}, and proceed to apply the diagrammatic method to a relativistic two-particle SR system in Section \ref{subsec:rel_SR}.

After building these foundations we turn in Section \ref{subsec:diag_density} to apply our revised diagrammatic method to the Hamiltonian introduced earlier in Section \ref{sec:covariant}. We finally solve the simple case of two particles travelling with offset relativistic velocities in Section \ref{sec:rel_sr_results}, where we establish a metric for SR enhancement as a function of velocity separation in the observer frame. We evaluate this metric for samples travelling at various highly relativistic speeds and we analyze the resulting velocity coherence requirements in the observer frame in the context of the standard relativistic transformation of velocity differentials.

\section{\label{sec:existing}Existing Models of Superradiance}

\citet{Dicke1954} first analyzed the so-called small sample limit wherein all particles are contained within a volume of dimensions much less than the emission wavelength. He considered the expectation values $\rho_m$ of populations of all states $|S\left(N,m\right)\rangle$ as evolving according to the transition rates $\left\{\mathcal{P}_{m,m-1}\right\}$. The resulting responses of the populations $\rho_m$ were used to determine the total intensity transient which peaked after some delay at a maximum emission rate greatly exceeding that predicted by spontaneous emission. Dicke continued in his foundational work to describe SR within an extended volume, to make realistic revisions to his model, and to analyze the effect of momentum back-reaction during photon emission.

Later models of SR approached the problem from either the master equation of quantum optics or the Heisenberg picture evolution of coarse grained inversion and polarization operators coupled to the quantized radiation field. These approaches lead to the Maxwell-Bloch equations briefly mentioned in Section \ref{sec:intro}. The Maxwell-Bloch equations are extremely useful for large samples which are intractable with purely quantum mechanical methods; however, they do not describe the initial quantum mechanical stage of SR. Moreover, our objective is to establish the most simple model possible for asking fundamental questions of a relativistic SR sample. The Maxwell-Bloch model is overly complicated for our present purposes.

We choose instead to generalize the small sample Dicke model \citep{Dicke1954} to a relativistic sample; however, as briefly discussed in Section \ref{sec:intro}, we deviate from his original methods. The Dicke model imagines the system evolving via discrete transitions through definite quantum states, essentially constructing a stochastic time sequence of events. Such an analysis intrinsically involves a non-relativistic observer and it is not at all clear how to translate this stochastic time sequence into a relativistic framework. Rather than attempt to analyze the role of relativistic effects in the cascade, we choose to apply the rote diagrammatic method of quantum optics to compute the fully quantum mechanical time evolution operator from our relativistically-derived Hamiltonian, and to introduce observer frame intensities only at the end of the analysis as measurements upon the formally evolved quantum state of the system.

\section{\label{sec:covariant}The Boussiakou, Bennett, and Babiker Relativistic Model of a Two-Level Particle}

There exists in the literature a quantum mechanical model of a two-level particle \citep{Boussiakou2002} derived from a relativistically covariant Lagrangian describing the matter-field interaction. Canonical quantization of the resulting observer frame Hamiltonian yields a Klein-Gordon-like description of the particle's center of mass, as well as relativistic $\gamma$ factor corrections to certain matter-field interaction terms. We refer to this Hamiltonian as the Boussiakou, Bennett, and Babiker (BBB) model of a relativistic two-level particle. We state the Hamiltonian below in this Section but leave its detailed synopsis to Appendix \ref{app:bbb_derivation}.

In a confirmation of the BBB model's relativistic foundations, it is shown in Section \ref{sec:rel_sr_theory} to properly describe the relativistic time dilation of the spontaneous decay of an initially excited particle, now in our diagrammatic framework (a recovery of the result already obtained in \cite{Boussiakou2002}). The BBB model is well suited to the task of constructing a relativistic SR model. We describe in Appendix \ref{app:bbb_derivation} its derivation at a high level as well as features and slight revisions pertinent to our work; for its detailed derivation, see \cite{Boussiakou2002}. The BBB Hamiltonian in the observer frame reads,
\begin{align}
    \hat{H} &= \hat{H}_a^0 + \hat{H}_f^0 + \hat{V} \\
    \hat{H}_a^0 &= \hat{H}_{\text{KG, com}} + \frac{1}{\gamma} \hbar \omega'_0 \hat{\pi}^\dagger \hat{\pi} \\
    \hat{H}^0_f &= \sum_{\bf k} \hbar \omega_{\bf k} \hat{a}^\dagger_{\bf k} \hat{a}_{\bf k} \\
    \begin{split}
        \hat{V} &= \sum_{\bf k} \left[ -\left(\frac{1}{\gamma}\vec{d}'_\parallel+\vec{d}'_\perp\right)\cdot \epsilon_{\bf k} + \left(\vec{d}'\times \epsilon^\perp_{\bf k}\right)\cdot \frac{{\bf V}_i}{c} \right] \\
        &\qquad \times \left(\xi_{\bf k} \hat{L}^\dagger_{\bf k} \hat{\pi}^\dagger \hat{a}_{\bf k} + \xi^*_{\bf k} \hat{L}_{\bf k} \hat{\pi} \hat{a}^\dagger_{\bf k} \right),
    \end{split}
\end{align}
where numerous terms require definition. The center of mass motion is described by the Klein-Gordon term
\begin{equation}
    \hat{H}_{\text{KG, com}} = \int \textrm{d}^3 P E_{\bf P}
    | {\bf P} \rangle \langle {\bf P}|,
\end{equation}
for ${\bf P} = \hbar {\bf K}$ the center of mass three-momentum of a Klein-Gordon particle of energy
\begin{equation}
E_{\bf P} = \sqrt{M^2 c^4 + \hbar^2 c^2 K^2} = \gamma_{\bf P} M c^2.
\end{equation}
The operators $\hat{\pi}^\dagger$ and $\hat{\pi}$ raise and lower the internal energy state of the particle, respectively.

Within the free field Hamiltonian $\hat{H}^0_f$ the operators $\hat{a}^\dagger_{\bf k}$ and $\hat{a}_{\bf k}$ create and annihilate (respectively) photons of mode ${\bf k}$ and linear polarization $\epsilon_{\bf k}$, where $\epsilon^\perp_{\bf k}=\left({\bf k} / k\right) \times\epsilon_{\bf k}$. The single-photon electric field coefficient is
\begin{equation}
    \xi_{\bf k} = i \sqrt{\frac{\hbar\omega_{\bf k}}{2 \epsilon_0 V_\textrm{Q}}}
\end{equation}
for $V_\textrm{Q}$ a fiducial field quantization volume and $\epsilon_0$ the permittivity of free space. We imply in all summations over field modes ${\bf k}$ a summation also over two orthogonal polarizations.

Within the interaction term $\hat{V}$ the operators $\hat{L}^\dagger_{\bf k}$ and $\hat{L}_{\bf k}$ raise and lower the Klein-Gordon center of mass momentum by $\hbar {\bf k}$, respectively; i.e.,
\begin{align}
    \hat{L}^\dagger_{\bf k} &= \int \mathrm{d}^3 P
    |{\bf P} + \hbar {\bf k}\rangle \langle {\bf P}| \label{eq:BBB_KG_mom_raise}\\
    \text{ and } \hat{L}_{\bf k} &= \int \mathrm{d}^3 P
    |{\bf P} - \hbar {\bf k}\rangle \langle {\bf P}|. \label{eq:BBB_KG_mom_lower}
\end{align}
The notation $\hat{L}_{\bf k}$ ($\hat{L}^\dagger_{\bf k}$) is unique to this work. The corresponding terms in \cite{Boussiakou2002} read $\exp\left(\pm i{\bf k} \cdot \hat{\bf Q}\right)$ for a center of mass position operator $\hat{\bf Q}$; i.e.,
\begin{equation}
    e^{\pm i{\bf k} \cdot \hat{\bf Q}} \left|{\bf P}\right\rangle = |{\bf P} \mp {\bf k} \hbar\rangle.
\end{equation}
For simplicity we have suppressed magnetization terms and restrict our work to the case of a particle possessing a rest frame electric dipole moment $\vec{d'}$. The dipole moment vector is composed of $\vec{d'_\parallel}$ and $\vec{d'_\perp}$ parallel and perpendicular to the particle's velocity, respectively.

\section{\label{sec:rel_sr_theory}Relativistic Two-Particle SR: Theory}

We develop in this section the diagrammatic representation of a relativistic two-particle SR sample. We assume familiarity with the diagrammatic representation of the SR time evolution operator; however, we provide a summary of the method in Appendix \ref{app:diagram_teo}.

The velocity dependent small sample SR problem is complicated by the localization of particles to within the wavelength of emission. Our development of the diagrammatic method therefore deviates from conventional methods, as we introduce in Section \ref{subsec:compromised_basis} a compromise between the position basis and the momentum basis which departs from a standard convolution structure \citep{Benedict1996}.

\subsection{\label{subsec:compromised_basis}Center of mass representation and localization considerations}

Seeking in this work a simple model of relativistic SR, we restrict our analysis to the small sample limit defined such that all particles are contained within a volume of dimensions much smaller than the wavelength of emission $\lambda$. The Heisenberg uncertainty relation then implies that $\Delta P_i \gg \hbar / \left(2 \lambda\right)$ in the initial momentum of any particle. This uncertainty could be realized, for example, by an initial Gaussian superposition in the center of mass Hilbert subspace. A particle initially excited with no photon in the radiation field could be described by the (unnormalized) state
\begin{equation}
    |\Psi\left(t=t_0\right)\rangle = \left[ \int \mathrm{d^3}P' e^{-\left({\bf P}'-{\bf P}_i\right)^2/\left(2\hbar\lambda\right)^2} |{\bf P}'\rangle\right] \otimes |e, 0\rangle. \label{eq:ex_mom_unc_super}
\end{equation}
Alternatively, the center of mass could be considered well-localized at some position ${\bf R}_i$ such that
\begin{equation}
    |\Psi\left(t=t_0\right)\rangle = |{\bf R}_i\rangle \otimes |e, 0\rangle.
\end{equation}

The literature treatment of diagrammatic SR \citep{Benedict1996} conventionally adopts the latter approach, working in the position basis and choosing to neglect the changes in momentum caused by the actions of $\hat{L}_{\bf k}$ or $\hat{L}^\dagger_{\bf k}$ in the interaction term $\hat{V}$. Alternatively, in a perfectly formal execution of such calculations, the initial state of the system would be defined by a construction similar to Eq. \eqref{eq:ex_mom_unc_super}, and modifications by $\hat{L}_{\bf k}$ and $\hat{L}^\dagger_{\bf k}$ to all ${\bf P}'$ in the superposition would be tracked during the system's time evolution. This exercise would greatly complicate calculations.

In our study of velocity dependent relativistic SR the initial particle momenta are critical to the velocity coherence analysis. We choose the following compromise between the strictly formal use of a superposition (such as Eq. \eqref{eq:ex_mom_unc_super}) and the literature approach of working in the position basis without tracking the particle momenta. We work in the momentum basis but we implicitly assume always that a center of mass momentum state $|{\bf P}\rangle$ is in fact a coarse distribution about a central value ${\bf P}$. The intuitive picture is that momentum states overlap with coarseness much greater than $\hbar / (2 \lambda)$.

In practice the coarse momentum assumption is important to transition matrix elements. Suppose two particles starting with momenta ${\bf P}_1$ and ${\bf P}_2$ evolve according to the BBB Hamiltonian. We omit their momenta in our description of the state of the system, but imply always that their values are consistent with conservation of momentum through transition matrix elements. For example, consider the matrix element $\tilde{V}_{g,{\bf k};1,0}\left(t\right)$ for a two-particle system, where the particle energy state $|1\rangle$ corresponds to only the first particle excited, $|g\rangle$ to both particles in the ground state, and $|0\rangle$/$|{\bf k}\rangle$ to the zero photon / single photon of mode ${\bf k}$ states. This matrix element is implied to denote a transition between any two coarse momentum states consistent with momentum conservation in their central values; i.e.,
\begin{equation}
    \tilde{V}_{g,{\bf k};1,0}\left(t\right) = \langle g, {\bf k}, {\bf P}_1 - \hbar {\bf k}, {\bf P}_2|\widehat{\tilde V}\left(t\right)|1, 0, {\bf P}_1, {\bf P}_2\rangle, \label{eq:trans_ex}
\end{equation}
which is related to the non-interaction picture matrix element by
\begin{equation}
    \tilde{V}_{g,{\bf k}; 1,0}\left(t\right) = e^{-i\left(\omega'_0/\gamma - \alpha_{{\bf k}1} \omega_{\bf k}\right)t} V_{g,{\bf k};1,0}\left(t\right), \label{eq:mom_cons}
\end{equation}
where $\alpha_{{\bf k}1}=1-\cos \theta_{{\bf k}1} V_1 / c$ for $\theta_{{\bf k}1}$ the angle between ${\bf P}_1$ and ${\bf k}$ (this result is obtained in the next section, viz. Eq. \eqref{eq:omega_change}). Whether momentum conservation is expressed by modifying the momentum in the bra or the ket of Eq. \eqref{eq:mom_cons} does not affect the results of later calculations.

The situation is further nuanced by considerations of the cascade history. It is argued in the literature \citep{Benedict1996} that the density operator formalism used in Section \ref{subsec:diag_density} below does not need to reference multi-photon states; however, when particle momenta are carefully tracked (which is not the case in \cite{Benedict1996}), prior photons emitted during the SR cascade affect later transition matrix elements. For example, suppose that the sample begins doubly excited and that particle 2 first emits a photon of mode ${\bf k}'$, such that its momentum is equal to ${\bf P}_2 - \hbar {\bf k}'$ prior to the transition process of Eq. \eqref{eq:trans_ex}. The resulting matrix element of Eq. \eqref{eq:mom_cons} would then be modified. In \cite{Benedict1996} the particle momenta are suppressed and such complications are avoided; this suggests that their methods are deficient for our present (relativistic) velocity dependent purposes, which necessarily reference the particle momenta.

We take the following approach. Instead of complicating the diagrammatic methods of the literature by tracking the particle momenta and generalising existing techniques to multi-photon states, we choose to suppress the particle momenta everywhere, save for transition matrix elements such as shown in Eq. \eqref{eq:mom_cons}. This is admittedly ad-hoc, but we are reassured by the success of the approach demonstrated in later sections. The model will be found to properly describe the continuum between the totally coherent limit where the particle velocity difference $\Delta v$ vanishes and the independent spontaneous emission limit $\Delta v \rightarrow \infty$; additionally, it will be found to accurately recover time dilation in the totally coherent case.

\subsection{\label{subsec:self_en_bbb}Convolution structure and the self-energy term under the BBB Hamiltonian}

We now establish matrix elements for the time evolution operator, beginning from its standard series expansion identities of Appendix \ref{app:diagram_teo}. As demonstrated in the literature \citep{Benedict1996}, Eqs. \eqref{eq:tev_o2} and \eqref{eq:tev_series} possess convolution structures. For example, the $U_{e,0;e,0}$ matrix element in the non-interaction picture may be expressed via Eq. \eqref{eq:tev_o2} as
\begin{align}
    U_{e,0;e,0}\left(t,t_0\right) &= F_{e,0}\left(t,t_0\right) \nonumber \\
    &-\frac{1}{\hbar^2} \sum_{\mathbf{k}} |f_{\bf k}|^2 \left[ F_{e,0} \star \left(F_{g,{\bf k}} \star U_{e,0;e,0}\right) \right] \left(t,t_0\right) \label{eq:Uee_convolution}
\end{align}
where $f_{\bf k}$ depends upon the initial velocity and dipole moment orientation of the particle according to
\begin{equation}
    f_{\bf k} = \xi_{\bf k} \left[ -\left(\frac{1}{\gamma}\vec{d}'_\parallel+\vec{d}'_\perp\right)\cdot \epsilon_{\bf k} + \left(\vec{d}'\times\epsilon^\perp_{\bf k}\right)\cdot \frac{{\bf V}_i}{c} \right].
\end{equation}
$F_{e,0}$ is the free theory propagator for a particle in the excited state and $F_{g,{\bf k}}$ the free theory propagator for a particle in the ground state with a single photon of mode ${\bf k}$ in the radiation field.

A Laplace transform replaces the convolution structure of Eq. \eqref{eq:Uee_convolution} with an algebraic relation; denoting the propagator Laplace transforms as $\mathcal{F}_{e,0}\left(s\right)$, $\mathcal{F}_{g,{\bf k}}\left(s\right)$, and $\mathcal{U}_{e,0;e,0}\left(s\right)$, we have the relation
\begin{equation}
    \mathcal{U}_{e,0;e,0} = \mathcal{F}_{e,0}\left[1 - \frac{1}{\hbar^2} \sum_{\bf k} |f_{\bf k}|^2 \mathcal{F}_{g,{\bf k}} \mathcal{U}_{e,0;e,0}\right]. \label{eq:prop_recur}
\end{equation}
In the interaction picture Eq. \eqref{eq:prop_recur} reads
\begin{equation}
    \tilde{\mathcal{U}}_{e,0;e,0} = \frac{1}{s} - \frac{1}{s}\frac{1}{\hbar^2} \sum_{\bf k} |f_{\bf k}|^2 \frac{1}{s+i\left(\omega_{g,{\bf k}}-\omega_{e,0}\right)} \tilde{\mathcal{U}}_{e,0;e,0}. \label{eq:prop_recur_int}
\end{equation}
The summation
\begin{equation}
    \Xi\left(s\right) = \frac{1}{\hbar^2} \sum_{\bf k} |f_{\bf k}|^2 \frac{1}{s+i\left(\omega_{g,{\bf k}}-\omega_{e,0}\right)} \label{eq:lap_self_energy}
\end{equation}
is conventionally referred to as the (Laplace domain) self-energy term, and it now deviates from the literature due to the velocity dependence as well as our use of the BBB Hamiltonian.

We require an expression for $\Xi$ for two different orientations of the dipole moment $\vec d'$ relative to the initial velocity of the particle. Recall that the mode identifier ${\bf k}$ is implied to extend over two polarization directions, which we now explicitly identify as
\begin{equation}
    \epsilon_{{\bf k}\theta} = \epsilon_{{\bf k}\phi} \times \frac{\bf k}{\left|\bf k\right|} \textrm{ and } \epsilon_{{\bf k}\phi} = \frac{\vec d' \times {\bf k}}{\left|\vec d' \times {\bf k}\right|}, \label{eq:pol_directions}
\end{equation}
with associated perpendiculars
\begin{equation}
    \epsilon^\perp_{{\bf k}\theta} = \epsilon_{{\bf k}\phi} \textrm{ and } \epsilon^\perp_{{\bf k}\phi} = -\epsilon_{{\bf k}\theta}
\end{equation}
by our earlier definition $\epsilon^\perp_{\bf k}=\left({\bf k} / |{\bf k}|\right) \times\epsilon_{\bf k}$. When ${\bf k}$ is parallel to ${\bf V}_i$ (and therefore also to $\vec d'$) Eq. \eqref{eq:pol_directions} becomes undefined, but by the symmetry of such a situation two orthogonal polarization directions may be arbitrarily assigned (in the plane perpendicular to ${\bf k}$) without affecting what follows.

From the BBB Hamiltonian we have that $\omega_{e,0}=\gamma M c^2 / \hbar + \omega'_0/\gamma$ and that
\begin{align}
    \omega_{g,{\bf k}} &= \sqrt{M^2 c^4 + \hbar^2 c^2 \left({\bf K}_i-{\bf k}\right)} + \omega_{\bf k} \nonumber \\
    &\approx \frac{\gamma M c^2}{\hbar} - {\bf V}_i \cdot {\bf k} + \omega_{\bf k}
\end{align}
for an initial center of mass velocity ${\bf V}_i$ (with which the $\gamma$ factor is identified). We thus have
\begin{equation}
    \omega_{e,0}-\omega_{g,{\bf k}} = \omega'_0/\gamma - \omega_{\bf k}\left[1-\frac{V_i}{c}\cos\left(\theta_{\bf k}\right)\right] \label{eq:omega_change}
\end{equation}
for an angle $\theta_{\bf k}$ between ${\bf k}$ and ${\bf V}_i$. Let us define
\begin{equation}
    \alpha_{\bf k} = 1-\beta_i \cos\left(\theta_{\bf k}\right)
\end{equation}
for $\beta_i = V_i / c$.

We consider first the case where ${\bf V}_i$ is parallel to $\vec d'$, in which case only the $\epsilon_{{\bf k}\theta}$ polarizations yield non-vanishing interaction matrix elements
\begin{equation}
    \langle e, 0| V |g, {\bf k}\theta\rangle = \frac{d'}{\gamma} \sin\left(\theta_{\bf k}\right) \xi_{\bf k},
\end{equation}
such that the self-energy term of Eq. \eqref{eq:lap_self_energy} becomes
\begin{equation}
    \Xi\left(s\right) = \frac{d'^2}{\hbar 2 \epsilon_0 \gamma^2 V_\textrm{Q}} \sum_{\bf k} \omega_{\bf k} \sin^2\left(\theta_{\bf k}\right) \frac{1}{s + i\left(\omega'_0/\gamma - \alpha_{\bf k} \omega_{\bf k}\right)}. \label{eq:self_en_step_a}
\end{equation}
Repeating the arguments of \cite{Benedict1996} (now for our relativistic Hamiltonian), in the absence of interaction equation \eqref{eq:prop_recur_int} possesses a pole at $s=0$. We therefore expect, in the perturbed system, that the neighbourhood of $s=0$ gives the main contribution to the sum over ${\bf k}$ such that we may approximate the self-energy factor of expression \eqref{eq:self_en_step_a} by evaluating it at $s=\epsilon$ and taking the limit $\epsilon \rightarrow 0$ \citep{Benedict1996}. Using the Sokhotski–Plemelj identity \cite{Henrici1986}
\begin{equation}
    \lim_{\epsilon \to 0^+} \frac{1}{i x + \epsilon} = \pi \delta\left(x\right) - i \textrm{Pr} \frac{1}{x},
\end{equation}
(where $\textrm{Pr}$ denotes the Cauchy principal value) and expressing the sum over ${\bf k}$ as an integral via
\begin{equation}
    \sum_{\bf k} \rightarrow \frac{V_\textrm{Q}}{(2\pi)^3} \int \mathrm{d}^3 k = \frac{V_\textrm{Q}}{(2\pi)^3} \int \mathrm{d}\Omega \int_0^\infty \mathrm{d} k k^2
\end{equation}
Eq. \eqref{eq:self_en_step_a} becomes
\begin{align}
    &\frac{d'^2}{\hbar \epsilon_0 (4 \pi \gamma)^2} \int \mathrm{d}\Omega \sin^2\left(\theta_{\bf k}\right) \int_0^{\infty} \mathrm{d}k k^2 \omega_{\bf k} \delta\left(\frac{\omega'_0}{\gamma} - \alpha_{\bf k} \omega_{\bf k}\right) \nonumber \\
    &=\frac{d'^2\omega'^3_0}{8 \pi \hbar \epsilon_0 c^3 \gamma^5} \int_0^\pi \mathrm{d}\theta_{\bf k} \frac{\sin^3\left(\theta_{\bf k}\right)}{\alpha_{\bf k}^4} \nonumber \\
    &=\frac{d'^2\omega'^3_0}{8 \pi \hbar \epsilon_0 c^3 \gamma^5} \int_{-1}^{1} \mathrm{d}x \frac{\left(1-x^2\right)}{\left[1-x \beta_i\right]^4} = \frac{\Gamma'_0}{2\gamma} \label{eq:Xi_result}
\end{align}
for $\Gamma'_0$ the rest frame spontaneous decay rate. In the time domain Eq. \eqref{eq:Xi_result} reads
\begin{equation}\Xi\left(t-t_0\right) = \frac{\Gamma'_0}{2\gamma} \delta \left(t-t_0\right). \label{eq:Xi_result_td}
\end{equation}
We have discarded the imaginary principal value part leading to the Lamb-like energy level shift, which we assume has been included in the definition of $\omega'_0$.

The fact that the self-energy $\Xi$ obtained by our diagrammatic approach corresponds to half the (time dilated) spontaneous emission rate is understandable if we now use $\Xi$ to solve for $\tilde{\mathcal{U}}_{e,0;e,0}$ in Eq. \eqref{eq:prop_recur_int} and obtain
\begin{equation}
    \tilde{\mathcal{U}}_{e,0;e,0}\left(s\right) = \frac{1}{s+\Xi} = \frac{1}{s+\Gamma'_0/\left(2\gamma\right)}
\end{equation}
which has inverse Laplace transform
\begin{equation}
    \tilde{U}_{e,0;e,0}\left(t,t_0\right) = e^{-\Delta t \Gamma'_0/\left(2\gamma\right)}. \label{eq:Uee_td}
\end{equation}
Since $|\tilde{U}_{e,0;e,0}\left(t,t_0\right)|^2$ corresponds to the probability of finding the particle excited after a time $\Delta t = t-t_0$, Eq. \eqref{eq:Uee_td} tells us that the particle decays with time constant $\Gamma'_0 / \gamma$, which is the same decay behaviour calculated in \cite{Boussiakou2002}. Our time evolution operator approach, necessary to our SR application, indeed recovers the results obtained by \citet{Boussiakou2002}. Note that this result is their demonstration that the BBB Hamiltonian successfully describes time dilation of spontaneous emission, which was the driving motive for our use of the BBB Hamiltonian in the present work.

It remains to yet evaluate $\Xi$ when ${\bf V}_i$ is perpendicular to $\vec d'$. The calculation is similar to the above work and ultimately leads again to $\Xi = \Gamma'_0/(2\gamma)$. This is not a new result, but rather a reproduction of the conclusion of \cite{Boussiakou2002}, unique only in our use of the time evolution operator and Laplace domain methods.

\subsection{\label{subsec:rel_SR}Relativistic two-particle SR propagators}

We now have at our disposal all of the tools necessary to compute the propagators of a relativistic two-particle SR sample. The methods of this section depart from the standard diagrammatic methodology of the literature, as we shortly demonstrate that the velocity dependent problem in our coarse momentum framework (necessary to the small sample SR problem) no longer possesses a convolution structure in its propagator expansions.

We denote the doubly excited particle state $|e\rangle$, the state of only the first particle excited $|1\rangle$, the state of only the second particle excited $|2\rangle$, and the doubly ground state $|g\rangle$. In this section we in fact compute only the zero photon propagators, which will prove sufficient for our density operator work in Section \ref{subsec:diag_density}; however, we do provide the relativistic propagators for single particle transitions into single photon states in Appendix \ref{app:photon_em_propagator}. The results of Appendix \ref{app:photon_em_propagator} can be used to compute multi-particle photon emission propagators where future research demands. Because no photons are present in this section, we will omit the radiation field state in our notation by writing, for example, $\tilde{U}_{e;e}$ instead of $\tilde{U}_{e,0;e,0}$.

\subsubsection{\label{subsubsec:rel_SR_Uee}Doubly excited propagator}

Let us first find the propagator for the two particles to both remain in the excited state. We obtain from the recurrence relation of Eq. \eqref{eq:tev_o2} the relation
\begin{widetext}
\begin{align}
    \tilde{U}_{e;e} \left(t,t_0\right) = 1 &+ \left[-\frac{i}{\hbar}\right]^2 \int_{t_0}^t \mathrm{d}t_2 \int_{t_0}^{t_2} \mathrm{d}t_1 \sum_{\bf k} \tilde{V}_{e;1,{\bf k}}\left(t_2\right) \tilde{V}_{1,{\bf k};e}\left(t_1\right) \tilde{U}_{e;e}\left(t_1,t_0\right) \nonumber \\
    &\qquad + \left[-\frac{i}{\hbar}\right]^2 \int_{t_0}^t \mathrm{d}t_2 \int_{t_0}^{t_2} \mathrm{d}t_1 \sum_{\bf k} \tilde{V}_{e;2,{\bf k}}\left(t_2\right) \tilde{V}_{2;\mathbf{k},e}\left(t_1\right) \tilde{U}_{e;e}\left(t_1,t_0\right). \label{eq:Uee_sr}
\end{align}
\end{widetext}
We would be repeating ourselves to calculate the self-energy parts within these convolutions: upon expanding the interaction matrix elements of Eq. \eqref{eq:Uee_sr} the central summations become the time domain equivalents to the calculation which followed Eq. \eqref{eq:self_en_step_a} and yielded $\Xi\left(t_2-t_1\right)=\left[\Gamma'_0/\left(2\gamma\right)\right]\delta\left(t_2-t_1\right)$. Eq. \eqref{eq:Uee_sr} thus becomes
\begin{equation}
    \tilde{U}_{e;e} \left(t,t_0\right) = 1 - \frac{\Gamma'_0}{2 \hbar^2}\left(\frac{1}{\gamma_1}+\frac{1}{\gamma_2}\right) \int_{t_0}^{t} \mathrm{d}t_1 \tilde{U}_{e;e}\left(t_1,t_0\right), \label{eq:Uee_sr_simp}
\end{equation}
which has solution
\begin{equation}
    \tilde{U}_{e;e}\left(t,t_0\right) = e^{-\Gamma'_0\left(1/\gamma_1+1/\gamma_2\right)\left(t-t_0\right)/2}
\end{equation}
for $\gamma_1$ and $\gamma_2$ the relativistic factors of particle 1 and 2, respectively. We can approximate
\begin{equation}
    \left(\frac{1}{\gamma_1}+\frac{1}{\gamma_2}\right) \approx \frac{2}{\gamma}
\end{equation}
for $\gamma$ the Lorentz factor of the average of the velocities, where the above approximation is correct to first order in $\Delta v /c$ for a velocity difference $\Delta {\bf v}$ between the two particles. We finally have that
\begin{equation}
    \tilde{U}_{e;e}\left(t,t_0\right) = e^{-\Gamma'_0\left(t-t_0\right)/\gamma}.
\end{equation}
Note that squaring the above result confirms that the doubly excited state again decays as expected by time dilation.

\subsubsection{\label{subsubsec:rel_SR_Ue1}Singly excited propagators}

We now calculate the propagator for the system to commence in the singly excited state of either particle and remain in that state at a later time. This situation differs from existing propagator calculations and results in the literature, as we shortly demonstrate a loss of convolution structure rooted in our use of the coarse momentum framework necessary to velocity dependent SR. We compute the propagator for the state $|1\rangle$ where only the first particle is excited; the propagator for $|2\rangle$ may be easily obtained by symmetry.

The recurrence relation of Eq. \eqref{eq:tev_o2} reads, for this matrix element,
\begin{widetext}
\begin{align}
    \tilde{U}_{1;1} \left(t,t_0\right) = 1 &+ \left[-\frac{i}{\hbar}\right]^2 \int_{t_0}^t \mathrm{d}t_2 \int_{t_0}^{t_2} \mathrm{d}t_1 \sum_{\bf k} \tilde{V}_{1,0;g,{\bf k}}\left(t_2\right) \tilde{V}_{g,{\bf k};1,0}\left(t_1\right) \tilde{U}_{1;1}\left(t_1,t_0\right) \nonumber \\
    &\qquad + \left[-\frac{i}{\hbar}\right]^2 \int_{t_0}^t \mathrm{d}t_2 \int_{t_0}^{t_2} \mathrm{d}t_1 \sum_{\bf k} \tilde{V}_{1,0;g,{\bf k}}\left(t_2\right) \tilde{V}_{g,\mathbf{k};2,0}\left(t_1\right) \tilde{U}_{2;1}\left(t_1,t_0\right). \label{eq:U11_sr}
\end{align}
\end{widetext}
The first integration term on the right side of Eq. \eqref{eq:U11_sr} describes the process of propagating from singly excited to singly excited, transitioning to the ground state, and transitioning back to the singly excited state. This term is familiar to us, becoming in the non-interaction picture a simple convolution structure containing the usual self-energy expression $\Xi\left(t_2-t_1\right)$. It simplifies to
\begin{equation}
    - \frac{\Gamma'_0}{2 \gamma} \int_{t_0}^{t} \mathrm{d}t_1 \tilde{U}_{1;1}\left(t_1,t_0\right). \label{eq:U11_sr_selfenterm}
\end{equation}

The second line of Eq. \eqref{eq:U11_sr} departs from the convolution structure. This line describes a process by which (i) the system transitions from the first particle singly excited state to the second particle singly excited state; (ii) the second particle emits a photon; and (iii) the first particle absorbs the photon. This physical picture of transferring a photon between the two particles suggests that this term will play a central role in SR velocity coherence.

It will greatly simplify what follows if we now choose a relative orientation between the initial velocities and dipole moments of the particles. Let us suppose that the two particles are travelling in the same direction with mean relativistic velocity ${\bf v}$ ($|{\bf v}| = v = \beta c$) and velocity difference $\Delta {\bf v} = {\bf v}_2 - {\bf v}_1$ ($|\Delta {\bf v}| = \Delta v$) as measured in the observer frame, and that their dipoles are oriented parallel to each other but perpendicular to their shared velocity direction. The two interaction terms within the central summation in the second line of Eq. \eqref{eq:U11_sr} then become
\begin{align}
    &e^{i\omega'_0 t_2 / \gamma} \xi^*_{\bf k} d' \left(\sin\theta_{\bf k} - \beta \cos\phi_{\bf k}\right) e^{-i \alpha_{{\bf k}1} \omega_{\bf k} t_2} \nonumber \\
    &\quad \times e^{i \alpha_{{\bf k}2} \omega_{\bf k} t_1} \xi_{\bf k} d' \left(\sin\theta_{\bf k} - \beta \cos\phi_{\bf k}\right) e^{-i\omega'_0 t_1 / \gamma} \label{eq:conv_viol}
\end{align}
where $\alpha_{{\bf k}1/2}$ are the $\alpha_{\bf k}$ factors for ${\bf v}_{1/2}$, respectively.

The central two exponentials of Eq. \eqref{eq:conv_viol} form the convolution violating term
\begin{equation}
    e^{-i\left(\alpha_{{\bf k}1}\omega_{\bf k}t_2 - \alpha_{{\bf k}2}\omega_{\bf k}t_1\right)}
\end{equation}
which cannot be expressed as a function of $t_2-t_1$. The occurrence of this term may be traced back to our use of the coarse momentum framework, wherein we chose not to evolve the distribution of momentum eigenstates describing a localized particle. Had we evolved such a distribution, we would have indeed retained a convolution structure between every adjacent occurrence of $V$ in the series representations of propagators; however, we would also have been required to track the center of mass momenta in our kets, which would have dramatically increased the number of matrix elements $\tilde{V}_{nm}$. Despite removing the convolution structure, then, our approach remains more efficient than describing the state of the particle's center of mass via a superposition such as shown, for example, in Eq. \eqref{eq:ex_mom_unc_super}.

\begin{figure}
    \centering
    \includegraphics[width=1.\columnwidth, trim=0cm 0cm 0cm 0cm, clip]{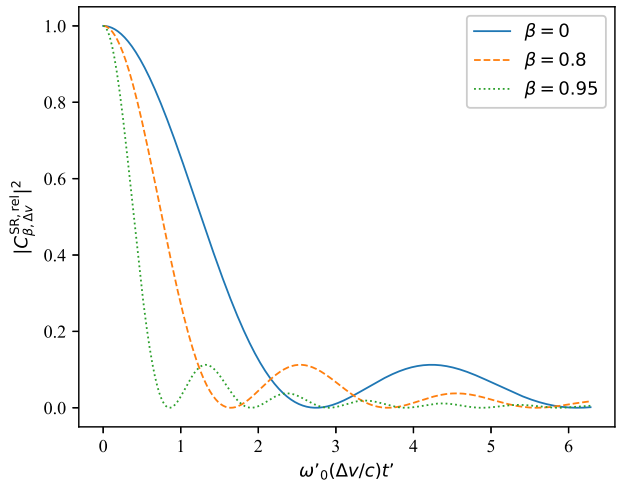}
    \caption{The squared norm of the relativistic SR velocity coherence kernel for $\beta=0,0.8,0.95$.}
    \label{fig:coh_kernel}
\end{figure}

We next evaluate the summation over ${\bf k}$ in the second line of Eq. \eqref{eq:U11_sr}. The details of this lengthy calculation are left to Appendix \ref{app:conv_viol_term}, where we find the entire second line of Eq. \eqref{eq:U11_sr} to be equal to
\begin{equation}
    -\frac{\Gamma'_0}{2 \gamma} \int_{t_0}^t \mathrm{d}t' C^\textrm{SR, rel}_{\beta, \Delta v}\left(t'\right) \tilde{U}_{2;1}\left(t',t_0\right),
\end{equation}
where
\begin{align}
    C^\textrm{SR, rel}_{\beta, \Delta v}\left(t'\right) = \frac{3}{8\gamma^2} \int_{-1}^{1} &\mathrm{d}x \frac{\left(1+\beta^2\right)\left(1+x^2\right)-4 \beta x}{\left[1-\beta x\right]^4} \nonumber \\
    &\times e^{i\omega'_0 \left(\Delta v / c\right) x t' / \left[\gamma \left(1-\beta x\right)\right]}.
\end{align}
We refer to $C^\textrm{SR, rel}_{\beta, \Delta v}$ as the relativistic SR velocity coherence kernel, which is an important tool in this work. A plot of $\left|C^\textrm{SR, rel}_{\beta, \Delta v}\right|^2$ is shown in Fig. \ref{fig:coh_kernel} for three values of $\beta$; we highlight features of $C^\textrm{SR, rel}_{\beta, \Delta v}$ at the end of this section after establishing its relationship to the two-particle propagators.

Eq. \eqref{eq:U11_sr} now reads
\begin{align}
    \tilde{U}_{1;1}\left(t,t_0\right) = 1 &- \frac{\Gamma'_0}{2\gamma} \int_{t_0}^{t} \mathrm{d}t_1 \tilde{U}_{1;1}\left(t_1,t_0\right) \nonumber \\
    &\quad -\frac{\Gamma'_0}{2 \gamma} \int_{t_0}^t \mathrm{d}t' C^\textrm{SR, rel}_{\beta, \Delta v}\left(t'\right) \tilde{U}_{2;1}\left(t',t_0\right),
\end{align}
which may be differentiated with respect to $t$ to obtain the first order differential equation
\begin{equation}
    \frac{\mathrm{d}}{\mathrm{d} t} \tilde{U}_{1;1} = -\frac{\Gamma'_0}{2\gamma} \tilde{U}_{1;1} - \frac{\Gamma'_0}{2\gamma} C^\textrm{SR, rel}_{\beta, \Delta v} \tilde{U}_{2;1}.
\end{equation}

Repeating all of the above procedure from Eq. \eqref{eq:U11_sr} onward for each of the possible single excitation propagators, we ultimately arrive at the set of coupled differential equations
\begin{align}
    \frac{\mathrm{d}}{\mathrm{d} t} \tilde{U}_{1;1} &= -\frac{\Gamma'_0}{2\gamma} \tilde{U}_{1;1} - \frac{\Gamma'_0}{2\gamma} C^\textrm{SR, rel}_{\beta, \Delta v} \tilde{U}_{2;1}\label{eq:prop_diffeq_1} \\
    \frac{\mathrm{d}}{\mathrm{d} t} \tilde{U}_{2;2} &= -\frac{\Gamma'_0}{2\gamma} \tilde{U}_{2;2} - \frac{\Gamma'_0}{2\gamma} C^\textrm{SR, rel}_{\beta, \Delta v} \tilde{U}_{1;2} \\
    \frac{\mathrm{d}}{\mathrm{d} t} \tilde{U}_{1;2} &= -\frac{\Gamma'_0}{2\gamma} \tilde{U}_{1;2} - \frac{\Gamma'_0}{2\gamma} C^\textrm{SR, rel}_{\beta, \Delta v} \tilde{U}_{2;2} \\
    \frac{\mathrm{d}}{\mathrm{d} t} \tilde{U}_{2;1} &= -\frac{\Gamma'_0}{2\gamma} \tilde{U}_{2;1} - \frac{\Gamma'_0}{2\gamma} C^\textrm{SR, rel}_{\beta, \Delta v} \tilde{U}_{1;1}. \label{eq:prop_diffeq_4}
\end{align}
After first numerically calculating the relativistic SR velocity coherence kernel, Eqs. \eqref{eq:prop_diffeq_1}–\eqref{eq:prop_diffeq_4} may be solved by any forward stepping numerical method from the initial conditions $\tilde{U}_{j;i}\left(t_0,t_0\right) = \delta_{ij}$. We use a fourth-order Runge Kutta stepping scheme for this purpose in later sections.

We point out the following reassuring properties of the relativistic SR velocity coherence kernel (see Fig. \ref{fig:coh_kernel}):
\begin{itemize}
    \item In the limit as $\Delta v \rightarrow \infty$ the kernel $C^\textrm{SR, rel}_{\beta, \Delta v} \rightarrow 0$; in this case Eqs. \eqref{eq:prop_diffeq_1}--\eqref{eq:prop_diffeq_4} are decoupled across the particles' identities such that they evolve independently.
    \item In the limit as $\Delta v \rightarrow 0$ the kernel $C^\textrm{SR, rel}_{\beta, \Delta v} \rightarrow 1$; this converts Eqs. \eqref{eq:prop_diffeq_1}--\eqref{eq:prop_diffeq_4} into the familiar coherent (zero velocity separation) propagator relations \citep{Benedict1996}.
    \item For non-relativistic $\beta$, $C^\textrm{SR, rel}_{\beta, \Delta v}$ has characteristic width $\omega'_0 (\Delta v / c) t' \approx 1$ (to order of magnitude); i.e., it yields SR enhancement between two particles over an SR timescale $T_\textrm{d}$ only if their Doppler separation is less than the characteristic frequency $1/T_\textrm{d}$ established by said timescale.
    \item The width of $C^\textrm{SR, rel}_{\beta, \Delta v}$ narrows with increasing $\beta$; i.e., velocity coherence requirements become stricter at increasingly relativistic velocities.
\end{itemize}

\subsection{\label{subsec:diag_density}Diagrammatic density operators for relativistic two-particle SR}

Having in our possession propagators for the relativistic two-particle SR sample, we are now able to describe the system's density operator evolution. In this section we parallel the steps followed by \citet{Benedict1996}, but modify the standard method as our velocity dependent relativistic SR model requires.

A sample commencing in the initial state $|\Psi\left(t_0\right)\rangle$ has a density operator $\hat{\rho}\left(t\right)=|\Psi\left(t\right)\rangle\langle\Psi\left(t\right)|$ which evolves according to
\begin{equation}
    \rho\left(t\right) = \hat{U}\left(t,t_0\right) \rho \left(t_0\right) \hat{U}^\dagger \left(t,t_0\right).
\end{equation}
We will be interested shortly in the diagonal elements $\rho_{m,m}\left(t\right)$. Consider for a moment the simpler case of the single photon diagonal matrix elements from a single-particle system evolving according to the BBB Hamiltonian's time evolution operator. Using the non-interaction picture version of Eq. \eqref{eq:tev_o1} to obtain the propagator
\begin{equation}
    U_{g,{\bf k};e,0} = -\frac{i}{\hbar} f^*_{\bf k} \left[ F_{g,{\bf k}} \star U_{e,0;e,0}\right] \left(t-t_0\right),
\end{equation}
we have that
\begin{align}
    \rho_{g,{\bf k};g,{\bf k}}\left(t\right) &= \frac{|f_{\bf k}|^2}{\hbar^2} \left[ F_{g,{\bf k}} \star U_{e,0;e,0}\right] \left(t-t_0\right) \nonumber \\ &\times \rho_{e,0;e,0}\left(t_0\right) \left[ F^*_{g,{\bf k}} \star U^*_{e,0;e,0}\right] \left(t-t_0\right).\label{eq:rho_gk_gk}
\end{align}
Eq. \eqref{eq:rho_gk_gk} is not a convolution between the first and second lines; however, we are interested not in the diagonal single photon elements themselves, but in the trace over them. By a standard result of the literature \citep{Benedict1996} the trace operation restores the convolution structure. We now recover this result in our present BBB Hamiltonian and see how it is modified.

The trace over Eq. \eqref{eq:rho_gk_gk} becomes
\begin{widetext}
\begin{align}
    \sum_{\bf k} &\frac{|f_{\bf k}|^2}{\hbar^2} \int_{t_0}^{t} \mathrm{d}t_1 \int_{t_0}^{t} \mathrm{d}t'_1 e^{i\alpha_{\bf k}\omega_{\bf k}\left(t_1-t_1'\right)} U_{e,0;e,0}\left(t_1,t_0\right) U^*_{e,0;e,0}\left(t'_1,t_0\right) \rho_{e,0;e,0}\left(t_0\right) \nonumber \\
    &\qquad = \int_{t_0}^{t} \mathrm{d}t_1 \int_{t_0}^{t} \mathrm{d}t'_1 e^{-i\omega'_0\left(t_1-t_1'\right)/\gamma} \Xi\left(t_1-t_1'\right) U_{e,0;e,0}\left(t_1,t_0\right)  U^*_{e,0;e,0}\left(t'_1,t_0\right) \rho_{e,0;e,0}\left(t_0\right) \nonumber \\
    &\qquad = 2 \int_{t_0}^{t} \mathrm{d}t_1 \int_{t_0}^{t_1} \mathrm{d}t'_1 \Xi\left(t_1-t_1'\right) \left| U_{e,0;e,0}\left(t'_1,t_0\right)\right|^2  \rho_{e,0;e,0}\left(t_0\right) \nonumber \\
    &\qquad = 2 \int_{t_0}^{t} \mathrm{d}t_1 \left[\Xi \star \left| U_{e,0;e,0}\right|^2 \right]\left(t_1, t_0\right) \rho_{e,0;e,0}\left(t_0\right) \nonumber \\
    &\qquad = \frac{\Gamma'_0}{\gamma}\int_{t_0}^{t} \mathrm{d}t_1 \left|U_{e,0;e,0}\right|^2\left(t_1,t_0\right) \rho_{e,0;e,0}\left(t_0\right) \nonumber \\
    &\qquad = \frac{\Gamma'_0}{\gamma} \left[\left|U_{e,0;e,0}\right|^2 \star \left|U_{g,0;g,0}\right|^2 \right]\left(t,t_0\right)\rho_{e,0;e,0}\left(t_0\right), \label{eq:vert_demofinal}
\end{align}
\end{widetext}
where we moved from the first line to the second by recognizing the presence of the time domain version of Eq. \eqref{eq:self_en_step_a} and using the result which followed it in Eq. \eqref{eq:Xi_result_td}; from the second line to the third by limiting to half the integration domain and exploiting the $t_1 \leftrightarrow t'_1$ symmetry, as well as recognizing that the integrand containing $\Xi\left(t_1-t'_1\right) = \delta\left(t_1-t'_1\right) \Gamma'_0 / (2\gamma)$ will vanish except where $t_1 = t'_1$; and from the fourth line to the fifth by exploiting the delta distribution in $\Xi$. Note that $U_{g,0;g,0}\left(t,t_0\right)=e^{-i E_{{\bf P}_i} \left(t-t_0\right)/\hbar}$ for $E_{{\bf P}_i}$ the kinetic energy of the Klein-Gordon particle possessing momentum ${\bf P}_i$.

\begin{figure}
    \centering
    \includegraphics[width=.94\columnwidth, trim=0cm 0cm 0cm 0cm, clip]{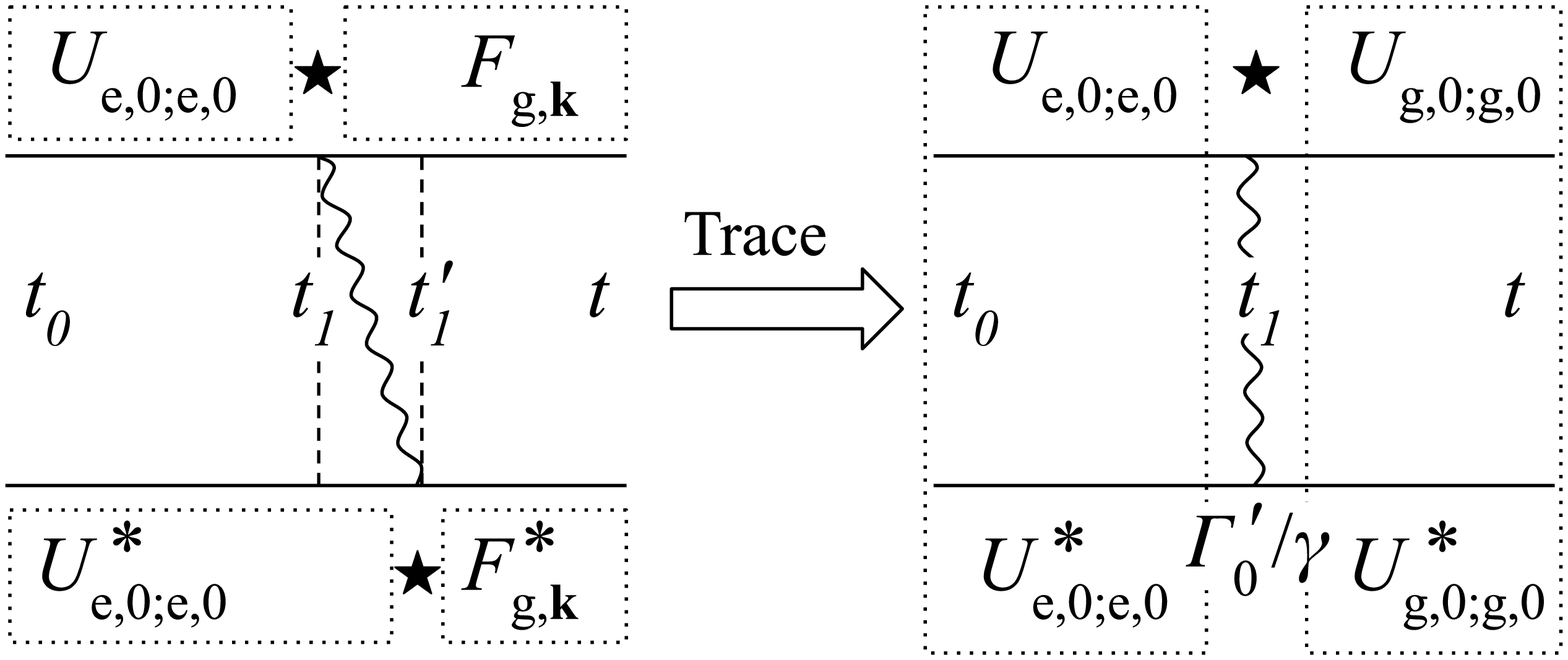}
    \caption{Diagrammatic representation of the standard vertical photon result of the literature for a single particle system which, in our present study of relativistic SR, now introduces a factor of $\Gamma'_0/\gamma$.}
    \label{fig:rho_diag_sp_em}
\end{figure}

Eq. \eqref{eq:vert_demofinal} tells us that the so-called vertical photon result is retained in the BBB Hamiltonian, but that it now introduces a factor of $\Gamma'_0/\gamma$. This result is critical to our density operator work to follow and is depicted pictorially in Fig. \ref{fig:rho_diag_sp_em}. On the top of the diagram are propagators appearing without conjugation in Eq. \eqref{eq:rho_gk_gk}, and on the bottom are propagators appearing with conjugation in Eq. \eqref{eq:rho_gk_gk}. The photon is depicted by a squiggly line. The left diagram depicts the original Eq. \eqref{eq:rho_gk_gk} prior to the trace operation and contains a convolution between top propagators multiplied by a convolution between bottom propagators. The right diagram depicts the trace result of Eq. \eqref{eq:vert_demofinal}, which transforms the structure into the multiplication of the leftmost top and bottom propagators, convolved with the multiplication of the rightmost top and bottom propagators, and introducing an overall factor of $\Gamma'_0/\gamma$. Note also that the rightmost propagators change from propagators with a photon in the field to no-photon propagators.

\begin{figure}
    \centering
    \includegraphics[width=.88\columnwidth, trim=0cm 0cm 0cm 0cm, clip]{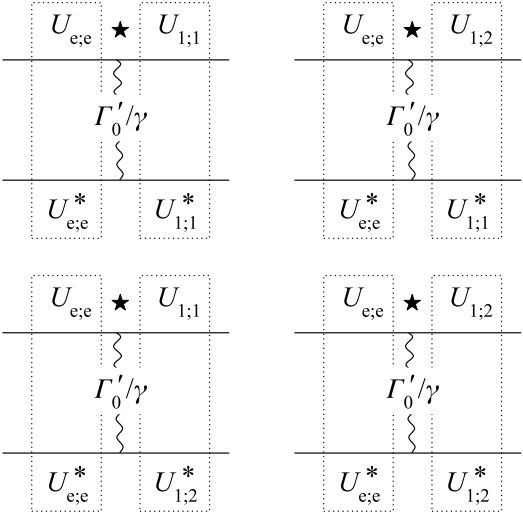}
    \caption{Diagrams corresponding to the quantity $\rho_{1;1}\left(t\right)$ for a two-particle relativistic SR sample starting in the doubly excited state $|e\rangle$. Note that time flows from left to right in each diagram, but that a propagator conventionally denotes transition from the state of its right index to the state of its left index.}
    \label{fig:rho11_diag_sr}
\end{figure}

Returning now to two-particle relativistic SR, the vertical photon result with accompanying $\Gamma'_0/\gamma$ factor is ubiquitous. Consider, for example, the quantity
\begin{equation}
    \rho_{1;1}\left(t\right) = \mathrm{Tr}_\mathrm{rad} \left\{ \langle 1,{\bf k}|\hat{\rho}\left(t\right)|1,{\bf k}\rangle\right\} = \sum_{\bf k} \rho_{1,{\bf k}; 1, {\bf k}} \left(t\right)
\end{equation}
which represents the probability of finding the system at time $t$ in the state with only the first particle excited. Similar calculations as preceding the single particle emission result of Eq. \eqref{eq:vert_demofinal} find that the four diagrams of Fig. \ref{fig:rho11_diag_sr} together describe the quantity $\rho_{1;1}\left(t\right)$. These four diagrams exhaust the possible permutations of propagators obeying the following constraints: first, the leftmost side of a diagram must commence in both its top and bottom propagators in the excited state, corresponding to the starting state of the system at $t=t_0$; second, transition across any vertical photon line introduces a factor of $\Gamma'_0 / \gamma$; third, in any region demarcated by vertical photon partitioning, propagators must couple states of like total number of excited particles, decreasing by one when crossing a vertical photon partition; and fourth, the singly excited propagators must terminate in the state $|1\rangle$ where the first particle in particular is excited, at least for this particular calculation of $\rho_{1;1}\left(t\right)$. Note when interpreting Fig. \ref{fig:rho11_diag_sr} that a propagator matrix element $U_{b;a}$ describes transition from $|a\rangle$ to $|b\rangle$; i.e., from right to left in its indices, even though time flows from left to right in all diagrams.

\begin{figure}
    \centering
    \includegraphics[width=.61\columnwidth, trim=0cm 0cm 0cm 0cm, clip]{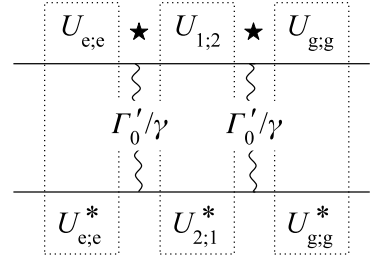}
    \caption{One of 16 possible diagrams corresponding to the quantity $\rho_{g;g}\left(t\right)$ for a two-particle relativistic SR sample starting in the doubly excited state $|e\rangle$.}
    \label{fig:rhogg_diag_sr}
\end{figure}

The diagrammatic rules outlined above generalize to more complicated diagrams describing the quantity $\rho_{g;g}\left(t\right)$. One such (randomly chosen) diagram is shown in Fig. \ref{fig:rhogg_diag_sr}. Propagators within the second (central) region of Fig. \ref{fig:rhogg_diag_sr} are no longer constrained to terminate in any particular state, which increases the number of total possible diagrams contributing to $\rho_{g;g}\left(t\right)$ to 16.

Actually, there is a great deal of symmetry to the set of differential Eqs. \eqref{eq:prop_diffeq_1}–\eqref{eq:prop_diffeq_4} and their initial conditions. From inspection, we see that $U_{1;2} = U_{2;1}$ and that $U_{1;1}=U_{2;2}$. Let us refer to the result of multiplying a top propagator with a bottom propagator within a region between vertical photons as a block propagator. In light of these symmetries and the constraints placed upon diagrams outlined above, the only possible block propagators for our system are
\begin{align}
    &B_{e} = \left|U_{e;e}\right|^2 \\
    &B_{a} = \left|U_{1;1}\right|^2, B_{b} = U^*_{1;1} U_{1;2}, B_{c} = \left|U_{1;2}\right|^2 \\
    &B_{g} = \left|U_{g;g}\right|^2 = 1,
\end{align}
as well as $B^*_{b}$.

We are finally prepared to assess velocity coherence in a relativistic two-particle SR sample possessing some mean $\beta$ and observer frame $\Delta v$, by the following steps:
\begin{enumerate}
    \item Solve the set of differential Eqs. \eqref{eq:prop_diffeq_1}–\eqref{eq:prop_diffeq_4} for the desired $\beta$ and $\Delta v$,
    \item Compute the diagram for each of $\rho_{1;1}\left(t\right)$, $\rho_{2;2}\left(t\right)$ and $\rho_{g;g}\left(t\right)$ as follows:
    \begin{enumerate}
        \item For each region (dotted rectangles in Figs. \ref{fig:rho11_diag_sr} and \ref{fig:rhogg_diag_sr}), compute the block propagator by multiplying the top propagator by the bottom propagator,
        \item Convolve all adjacent block propagators, and
        \item Multiply the result by a factor of $\Gamma'_0 / \gamma$ for each vertical photon in the diagram
    \end{enumerate}
    \item Sum all diagram results for a given $\rho_{a;a}$,
    \item In order to interpret a total emission power from the particle excitation time dependence, compute the time derivative of the total particle energy in the observer frame; i.e., $\left(2 \dot\rho_{e;e} + \dot\rho_{1;1} + \dot\rho_{2;2}\right) \hbar \omega'_0 / \gamma$. \label{last_step}
\end{enumerate}

In terms of our block propagators and symmetries, we have that
\begin{align}
    \rho_{e;e}\left(t\right) &= B_{e}\left(t\right) \\
    \rho_{1;1}\left(t\right) &= \frac{\Gamma'_0}{\gamma} \left\{ B_{e} \star \left[B_{a} + 2 \textrm{Re}\left(B_{b}\right) + B_{c}\right]\right\} \left(t\right) \\
    \rho_{2;2}\left(t\right) &= \rho_{1;1}\left(t\right) \\
    \rho_{g;g}\left(t\right) &= 4 \left(\frac{\Gamma'_0}{\gamma}\right)^2 \left\{ B_{e} \star \left[B_{a} + 2 \textrm{Re}\left(B_{b}\right) + B_{c} \right] \star B_{g}\right\} \left(t\right).
\end{align}
The above results are differentiated in step \ref{last_step} above, which allows us to skip one convolution from any of the above calculations; for example,
\begin{align}
    \dot\rho_{g;g}\left(t\right) &= 4 \left(\frac{\Gamma'_0}{\gamma}\right)^2 B_{e}\left(t\right) \left\{\left[B_{a} + 2 \textrm{Re}\left(B_{b}\right) + B_{c} \right] \star B_{g}\right\} \left(t\right) \nonumber \\
    &= 4 \left(\frac{\Gamma'_0}{\gamma}\right)^2 e^{-2\Gamma'_0\left(t-t_0\right)/\gamma} \nonumber \\
    &{\hskip4em\relax} \times \left\{\left[B_{a} + 2 \textrm{Re} \left(B_{b}\right) + B_{c} \right] \star B_{g}\right\}\left(t\right).
\end{align}

\section{\label{sec:rel_sr_results}Relativistic Two-Particle SR: Results}

The first result of our theory is a confirmation of the time dilation of two-particle SR in the $\Delta v \rightarrow 0$ limit. In this limit the relativistic coherence kernel goes to 1, so that all factors of $\gamma$ in Eqs. \eqref{eq:prop_diffeq_1}–\eqref{eq:prop_diffeq_4} may be removed by transforming to the variable $\tau = t/\gamma$. Whatever the result of performing all steps required to compute a total emission power transient, then, that result will be a function of $t/\gamma$ and thus time-contracted in the observer frame. This is not so much a useful result as it is a reassurance that our model behaves as expected.

The quantity of interest to this work is the behaviour of SR coherent enhancement as a function of the velocity difference of highly relativistic particles. Let us work in the variable $q = \Gamma'_0 t / \left(2\gamma\right)$ such that the only non symmetry-redundant quantities of Eqs. \eqref{eq:prop_diffeq_1}–\eqref{eq:prop_diffeq_4} are related by
\begin{align}
    \frac{\mathrm{d}}{\mathrm{d}q}{\tilde{U}}_{1;1} &= - \tilde{U}_{1;1} - C^\textrm{SR, rel}_{\beta, \Delta v} \tilde{U}_{2;1} \label{eq:U11_prop_qvar} \\
    \frac{\mathrm{d}}{\mathrm{d}q}{\tilde{U}}_{1;2} &= - \tilde{U}_{1;2} - C^\textrm{SR, rel}_{\beta, \Delta v} \tilde{U}_{1;1} \label{eq:U12_prop_qvar}.
\end{align}
We have also at the outset that $\tilde{U}_{e;e}\left(q\right) = e^{-2q}$.

A \texttt{Python} script was written to compute $C^\textrm{SR, rel}_{\beta, \Delta v}$, to solve Eqs. \eqref{eq:U11_prop_qvar} and \eqref{eq:U12_prop_qvar}, to compute the block propagators, to convolve the appropriate diagrams for the density operator traces, and finally to obtain the total power transient for various particle velocity differences and relativistic $\beta$ values.

\begin{figure}
    \centering
    \includegraphics[width=1.\columnwidth, trim=0cm 0cm 0cm 0cm, clip]{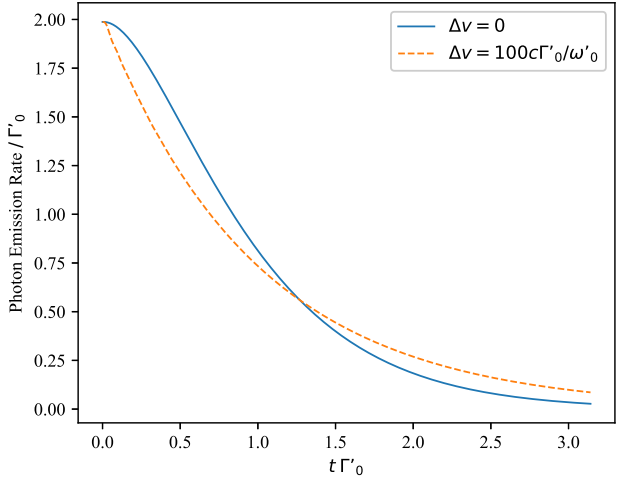}
    \caption{Photon emission rate as a function of time for two velocity separations $\Delta v = 0$ and $\Delta v = 100 c \Gamma'_0 / \omega'_0$ for $\beta=0$.}
    \label{fig:ind_demo}
\end{figure}

In the simplest case, we first verify that our model demonstrates a transition of SR enhancement between the far separated $\Delta v = \infty$ limit and the coherent $\Delta v = 0$ limit. The photon emission rate at $\beta=0$ is shown in Fig. \ref{fig:ind_demo} for the coherent case $\Delta v = 0$ and for the well separated case $\Delta v = 100 c \Gamma'_0 / \omega'_0$. The former demonstrates SR enhancement, while the latter recovers the spontaneous emission rate from two independent particles.

Fig. \ref{fig:ind_demo} does not differ substantially between the coherent SR ($\Delta v=0$) and the independent spontaneous emission ($\Delta v \rightarrow \infty$) cases. In order to obtain a more dramatic SR pulse demonstrating enhanced and delayed peak intensity one would need to simulate a larger number of particles. Such an exercise becomes very complicated very quickly, as the number of diagrams requiring evaluation grows extremely fast with the number of particles. Combinatoric considerations or computer methods could speed the process up and make a slightly larger population manageable, but we leave such exercises to future work. Instead, we focus now on extracting meaningful information from the two-particle case.

\begin{figure}
    \centering
    \includegraphics[width=1.\columnwidth, trim=0.4cm 0.4cm 0.1cm 0cm, clip]{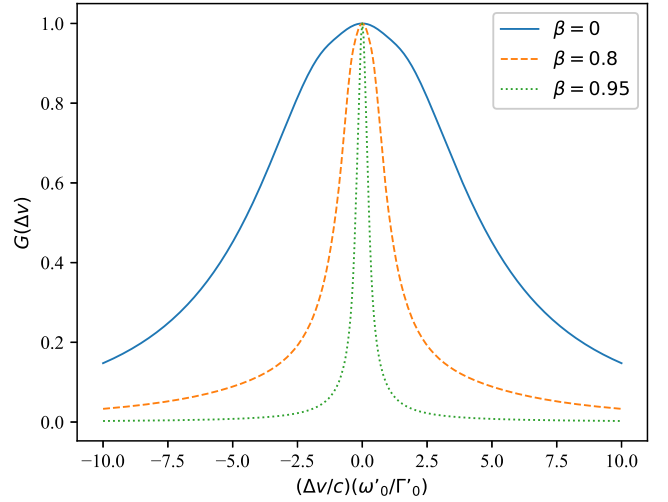}
    \caption{The relativistic velocity coherence metric $G\left(\Delta v\right)$ for $\beta=0,0.8,0.95$.}
    \label{fig:coh_analysis}
\end{figure}

For the purpose of quantifying the departure of an SR system from the independent limit as a function of observer frame velocity difference, let
\begin{equation}
    I_{\Delta v}\left(t\right) = 2 \dot\rho_{e;e} + \dot \rho_{1;1} + \dot \rho_{2;2}
\end{equation}
for a sample having particle velocity difference $\Delta v$, and define the relativistic velocity coherence metric
\begin{equation}
    G\left(\Delta v\right) = \frac{1}{A}\lim_{T\rightarrow\infty} \int_0^T \mathrm{d}t \left[I_{\Delta v}\left(t\right)-2\frac{\Gamma'_0}{\gamma}e^{-\Gamma'_0 t / \gamma} \right]^2
\end{equation}
which measures total departure of a given power transient from that transient which would be generated by independently emitting particles. The $1/A$ out front is a normalisation factor such that $G\left(0\right)=1$. Fig. \ref{fig:coh_analysis} shows $G\left(\Delta v\right)$ for three values of $\beta$. The full-width half maxima (FWHM) of the various $G\left(\Delta v\right)$ provide us with a measure of the velocity coherence required for two relativistic particles to demonstrate enhanced SR emission. For $\beta=0$ the FWHM is approximately $9.1 c \Gamma'_0 / \omega'_0$, while for $\beta=0.95$ the FWHM is approximately $0.52 c \Gamma'_0 / \omega'_0$.

There are numerous ways to interpret the various FWHM. In the simplest approach, let us compare their values in the context of the relativistic transformation of velocity differentials. The usual relativistic velocity composition rules state that small velocity differences transform as $\Delta v' \approx \gamma^2 \Delta v$ between the observer and rest frames. The $\beta=0.95$ plot of Fig. \ref{fig:coh_analysis} corresponds to $\gamma^2 \approx 10.3$; however, that plot's FWHM differed from the $\beta=0$ case by a factor of $17.5$. This suggests that the velocity coherence required between two particles to demonstrate enhanced SR emission in the observer frame is more restrictive than accounted for by simple considerations of the relativistic transformation of velocity spreads.

The situation is further nuanced if we ask which velocity scale should be assigned meaning when quantifying velocity coherence requirements. We have thus far compared velocity differences in terms of the relativistically invariant unit $c \Gamma'_0 / \omega'_0$; however, the researcher may be more interested in  absolute velocity differences in a selected reference frame. Whether velocity coherence requirements become more or less restrictive from a relativistic source is thus not well-defined in a general sense, and will depend upon what other interactions are of interest to the researcher. Moreover, many other relativistic effects may enter into the picture. We discuss some of these effects in Section \ref{sec:summary}.

\section{\label{sec:summary}Summary and Future Work}

We constructed a model of velocity dependent relativistic two-particle SR based upon an existing model \citep{Boussiakou2002} of relativistic spontaneous emission from a two-level particle. We applied the diagrammatic method of quantum optics to our relativistic SR Hamiltonian and found that certain standard results and techniques of the literature required revision. The introduction of velocity dependence caused a loss of convolution structure in zero-photon propagator calculations. We introduced a set of coupled differential equations, as well as the so-called relativistic velocity coherence kernel, which together enabled the calculation of SR propagators. The standard vertical photon result of the literature was re-proven and was shown to introduce an overall Lorentz factor into the density operator diagrams of SR.

We solved our relativistic SR model in Section \ref{sec:rel_sr_results} for two particles and characterized velocity coherence requirements for SR emission enhancement from relativistic sources at various $\beta$ factors. Initial results suggested that velocity coherence requirements for enhanced SR emission in the observer frame become more restrictive as a source moves with higher relativistic velocity. Some caveats were made in Section \ref{sec:rel_sr_results} regarding the interpretation of velocity coherence requirements, which in reality becomes a nuanced exercise.

The model constructed in this paper forms a foundation for relativistic SR work. Future research could proceed in numerous directions. First, it remains to yet solve the present model for (i) dipoles aligned parallel to the particles' velocities, (ii) dipoles not oriented parallel to each other, and (iii) non co-linear particle velocities. Second, the diagrammatic method may also be used to calculate the spectrum and angular photon distribution from an SR source. This procedure could be applied to the present model and radiation intensity could be characterized across different directions from a relativistic SR source. Some initial calculation steps for this purpose are provided in Appendix \ref{app:photon_em_propagator}. Third, this model could be expanded to a larger number of particles. Although the diagrammatic method becomes quickly cumbersome with increasing population, computer methods exploiting combinatoric considerations could certainly simulate a moderately sized sample. Despite remaining physically negligible, a sample of four or more particles begins to demonstrate the salient SR features of delayed and enhanced peak intensity. A relativistic model of these features would be especially helpful to astrophysicists. Fourth, this model could be generalized to an extended sample by paralleling existing derivations of the Maxwell-Bloch equations, now with our relativistic SR Hamiltonian. Considerations of relativistic length contraction or time dilation of relaxation processes in such an extended model could lead to richer relativistic SR behavior.

\begin{acknowledgments}
C.W. is supported by the Natural Sciences and Engineering Research Council of Canada (NSERC) through the doctoral postgraduate scholarship (PGS D). F.R.’s research at Perimeter Institute is supported in part by the Government of Canada through the Department of Innovation, Science and Economic Development Canada and by the Province of Ontario through the Ministry of Economic Development, Job Creation and Trade. M.H.'s research is funded through the Natural Sciences and Engineering Research Council of Canada Discovery Grant RGPIN-2016-04460 and the Western Strategic Support for NSERC Success Accelerator program.

The authors wish to thank Mr. M. Chamma and Ms. A. Kumar for their support through helpful conversations, proofreading, and general feedback.

\end{acknowledgments}

\appendix

\section{\label{app:bbb_derivation}Synopsis of the Original BBB Hamiltonian Derivation \cite{Boussiakou2002} and Manipulation into its Present Form}

In this Appendix we provide a synopsis of the physical motivation, derivation, and manipulation of the BBB Hamiltonian first presented by \citet{Boussiakou2002} and slightly revised in form for the purpose of this work.

\subsection{\label{subsec:BBB_hamiltonian}The BBB Hamiltonian}

Prior to canonical quantization, construction of the BBB model starts from the usual Lagrangian of a positive and a negative charge, with generalized coordinates corresponding to the center of mass coordinate and to the charge separation in a frame co-moving with the particle's center of mass. \citet{Boussiakou2002} select the Power-Zienau-Woolley (PZW) gauge for the electromagnetic field, which permits them to express the Lagrangian in the particle rest frame in a manifestly covariant form by contracting the field tensor with itself (the usual free field contribution to the Lagrangian) and contracting the field tensor with the polarization-magnetization tensor (a covariant description of the matter-field interaction). The polarization and magnetization fields comprising the latter tensor in the PZW gauge are constructed from the charge separation coordinate and its velocity via a prescription provided by existing PZW gauge theory in the literature \footnote{namely, Eqs. (6) and (7) of \cite{Boussiakou2002} for reference}.

The covariant Lagrangian formulation enables Boussiakou \textit{et al.} to determine, via careful Lorentz transformations, an expression for the observer frame Hamiltonian in terms of the observer frame center of mass coordinates, charge separation coordinates, and radiation field; these observer frame quantities are coupled to the particle's rest frame polarization and magnetization properties. This form of the Hamiltonian ultimately enables the BBB model to relate observer frame behaviour to transition operators intrinsic to the particle. The BBB Hamiltonian in the observer frame reads,
\begin{align}
    \hat{H} &= \hat{H}_a^0 + \hat{H}_f^0 + \hat{V} \\
    \hat{H}_a^0 &= \hat{H}_{\text{KG, com}} + \frac{1}{2\gamma\mu}\left(\hat{p}^2_\perp + \frac{\hat{p}^2_\parallel}{\gamma^2}\right) - \frac{e^2}{4\pi\epsilon_0\gamma \hat{q}'} \label{eq:BBB_Ha} \\
    \begin{split}
        \hat{V} &= \int \mathrm{d}^3r \left\{\frac{1}{\epsilon_0}\left[\hat{\bm{\mathcal{P}}}'_\parallel\left({\bf r}\right)+\gamma\hat{\bm{\mathcal{P}}}'_\perp\left({\bf r}\right)\right]\cdot\hat{{\bf E}}\left({\bf r}\right) \right. \label{eq:BBB_Hint} \\
        &\! + \left. \frac{{\bf \hat{P}}}{2M}\cdot\left[\hat{\bm{\mathcal{P}}}'\left({\bf r}\right)\times\hat{{\bf B}}\left({\bf r}\right)\right]+\left[\hat{\bm{\mathcal{P}}}'\left({\bf r}\right)\times\hat{{\bf B}}\left({\bf r}\right)\right]\cdot\frac{\hat{{\bf P}}}{2M}\right\},
    \end{split}
\end{align}
where a few outstanding terms have not yet been defined within the body of the paper. The position operators $\hat{q}_\parallel$ and $\hat{q}_\perp$ describe the charge separation coordinates (unprimed in the observer frame, primed in the rest frame) parallel and perpendicular to the center of mass velocity, the operators $\hat{p}_\parallel$ and $\hat{p}_\perp$ are their conjugate momenta operators, and the center of mass has an observer frame momentum operator $\hat{\bf P}$. The Lorentz factor $\gamma$ corresponds to the initial center of mass momentum of the particle. The operator $\hat{q}'$ is the charge separation distance operator in the rest frame coordinates, $\mu$ is the rest frame reduced mass of the particle, and $\hat{\bm{\mathcal{P}}}'$ is the particle's rest frame polarization operator. For simplicity we have suppressed magnetization terms, consistent with the paper's restriction to the case of an electric dipole which couples to the observer frame electric ($\hat{\bf E}$) and magnetic ($\hat{\bf B}$) fields. The second line of Eq. \eqref{eq:BBB_Hint} is the so-called R\"{o}ntgen interaction term.

\subsection{\label{subsec:BBB_zeroth}Eigenfunctions of the unperturbed BBB Hamiltonian}

We require the unperturbed eigenfunctions of the atomic Hamiltonian term defined by Eq. \eqref{eq:BBB_Ha}. We present here the main results of the solution provided in \cite{Boussiakou2002}; the eigenfunction work below is adapted directly from their work.

Projecting onto the observer frame center of mass and charge separation coordinate basis and factoring the wavefunction as $\Psi\left({\bf Q},{\bf q}\right)=e^{i{\bf K}\cdot{\bf Q}}\psi\left({\bf q}\right)$, one obtains
\begin{equation}
    \left[-\frac{1}{2\mu}\left(\nabla^2_{q,\perp}+\frac{1}{\gamma^2} \frac{\partial^2}{\partial q^2_\parallel} \right) - \frac{e^2}{4\pi\epsilon_0 q'} \right] \psi\left({\bf q}\right) = \gamma \epsilon \psi\left({\bf q}\right) \label{eq:BBB_schr}
\end{equation}
for $\epsilon$ the (observer frame) energy of the state. Eq. \eqref{eq:BBB_schr} may be expressed in rest frame coordinates as
\begin{equation}
    \left[-\frac{1}{2\mu} \nabla^2_{q'} - \frac{e^2}{4\pi\epsilon_0 q'} \right] \psi\left({\bf q}'\right) = \gamma \epsilon \psi\left({\bf q}'\right), \label{eq:BBB_schr_prime}
\end{equation}
which is the usual hydrogenic Schr\"{o}dinger equation, but where the energy eigenvalues of the conventional solutions correspond now to $\gamma$ times the energy eigenvalues in the observer frame. Additionally, reversion back to the observer frame will introduce a length contraction to the conventional hydrogenic wavefunction shapes. Both of these results \citep{Boussiakou2002} are consistent with familiar results from the special theory of relativity \citep{Einstein1905}.

It is important to keep in mind that the primed coordinates are simply a mathematical convenience for working with a wavefunction which is physically relevant to the observer frame. Although the unprimed and primed coordinates are related by the familiar Lorentz transformation between the observer and rest frame position coordinates, the primed coordinates serve only as a tool for solving Eq. \eqref{eq:BBB_schr}. The physics thus far developed has been based upon canonical quantization in the observer frame and meaningful conclusions from the present wavefunction can only be made with respect to that frame.

We now make an important point regarding normalization which affects our diagrammatic calculations in this work. The energy eigenfunction relation of Eq. \eqref{eq:BBB_schr} becomes hydrogenic in the primed coordinates of \eqref{eq:BBB_schr_prime}; we choose therefore to use the conventional hydrogenic wavefunctions in the primed coordinates, or their transformed version in the unprimed coordinates. Such functions are no longer normalized in the unprimed (observer) frame due to the factor of $\gamma$ relating the coordinates parallel to the particle's motion. We might alternatively re-normalize the eigenfunctions in the observer frame; however, we will later be interested in expanding operators in basis states indexed by the primed (rest frame) coordinates. We therefore retain the conventional normalization of the hydrogenic wavefunctions relative to the primed coordinates, but remembering always that the completeness must then be asserted relative to the primed basis states; i.e.,
\begin{equation}
    \hat{1} = \int \mathrm{d}^3 q' |{\bf q'}\rangle\langle{\bf q'}|.
\end{equation}

\subsection{\label{subsec:Hint_restdip}The BBB Hamiltonian in its form useful to this work}

In this section we summarize the work of \cite{Boussiakou2002} to express the BBB Hamiltonian in terms of operators and properties intrinsic to the particle, including internal raising and lowering operators $\hat{\pi}^\dagger$ and $\hat{\pi}$, the internal dipole moment $\vec{d}'$, and the rest frame transition energy $\hbar \omega'_0$. We also introduce a simplified form of the interaction Hamiltonian useful to this work.

The operator $\hat{\bm{\mathcal{P}}}'$ appearing in the interaction operator $\hat V$ of Eq. \eqref{eq:BBB_Hint} refers to the polarization vector operator in the rest frame, and it couples to field operators in the observer frame. \citet{Boussiakou2002} assert the dipole approximation to express the rest frame polarization operator as $\hat{\bm{\mathcal{P}}}'\left(\bf{r}'\right)= \hat{\bf{d}}'\delta\left({\bf r'- Q'}\right)$, where $\hat{\bf d}' = e \hat{\bf q}'$. With this approximation, we perform first the integration over $\mathrm{d}^3 r$ of Eq. \eqref{eq:BBB_Hint} in the observer frame; however, we must recognize that the arguments of the delta distribution are in the rest frame coordinates, and thus its support is length-contracted in the observer frame; i.e., the integration picks up a factor of $1/\gamma$ and we are left with
\begin{align}
    \begin{split}
        \hat{V} &= -\left(\frac{1}{\gamma}\hat{\bf d}'_\parallel+\hat{\bf d}'_\perp\right)\cdot \hat{\bf E}\left({\bf Q}\right) \\
        &\quad + \frac{\hat{\bf P}}{2M \gamma}\cdot\left[\hat{\bf d}'\times\hat{\bf B}\left({\bf Q}\right)\right]+\left[\hat{\bf d}'\times\hat{\bf B}\left({\bf Q}\right)\right]\cdot\frac{\hat{\bf P}}{2M \gamma}.
    \end{split}
\end{align}

Let us prepare ourselves for the theory of dipole emission between two energy eigenstates $|E_{1,2}\rangle$ of the particle. We expand the dipole operator as
\begin{align}
    \hat{\bf d}' &= e \hat{\bf q}' \nonumber \\
    &= e \sum_{m,n} |E_m \rangle \langle E_m|\hat{\bf q}'|E_n \rangle \langle E_n|.
\end{align}
The matrix elements are computed as
\begin{align}
    \langle E_m|\hat{\bf q}'|E_n \rangle &= \langle E_m|\hat{\bf q}'\int \mathrm{d}^3 \bar{q}' |\bar{\bf q}'\rangle \langle \bar{\bf q}'|E_n \rangle \nonumber \\
    &= \int \mathrm{d}^3 \bar{q}' \psi^*_m\left(\bar{\bf q}'\right) \bar{\bf q}'\psi_n\left(\bar{\bf q}'\right),
\end{align}
where we intentionally introduced the completeness relation over the primed coordinates as per our earlier discussion. This final expression is mathematically the same as the conventional dipole matrix element calculation for a stationary particle, having nonzero elements only when $m \neq n$. We need not perform the actual integration over the hydrogenic wavefunctions, since we are concerned only with relating our theory to that of a stationary particle; we thus have
\begin{align}
    \hat{\bf d}' = \vec{d}'\left(\hat{\pi}^\dagger + \hat{\pi}\right)
\end{align}
for $\hat{\pi}^\dagger$ and $\hat{\pi}$ the internal raising and lowering operators of the particle, and where $\vec{d}'$ is the rest frame dipole moment.

Summarizing the above results and expanding the field operators in terms of creation and annihilation operators, the BBB Hamiltonian is expressed using rest frame quantities and operators as
\begin{align}
    \hat{H}^0_a &= \hat{H}_{\text{KG, com}} + \frac{1}{\gamma} \hbar \omega'_0 \hat{\pi}^\dagger \hat{\pi} \label{eq:BBB_Ha_intrinsic} \\
    \begin{split}
        \hat{V} &= \sum_{\bf k} \left[ -\left(\frac{1}{\gamma}\vec{d}'_\parallel+\vec{d}'_\perp\right)\cdot \epsilon_{\bf k} \left( \xi_{\bf k} \hat{L}^\dagger_{\bf k} \hat{\pi}^\dagger \hat{a}_{\bf k} + \xi^*_{\bf k} \hat{L}_{\bf k} \hat{\pi} \hat{a}^\dagger_{\bf k} \right) \right. \\
        &+ \left. \frac{1}{2M\gamma c} \left(\vec{d}'\times\epsilon^\perp_{\bf k}\right)\cdot \left\{\hat{\bf P}, \xi_{\bf k} \hat{L}^\dagger_{\bf k} \hat{\pi}^\dagger \hat{a}_{\bf k} + \xi^*_{\bf k} \hat{L}_{\bf k} \hat{\pi} \hat{a}^\dagger_{\bf k} \right\}\right]. \label{eq:BBB_Hint_intrinsic}
    \end{split}
\end{align}

In this work we consider the evolution of an initially excited particle starting with a center of mass momentum ${\bf P}_i$. This initial state is coupled by Eq. \eqref{eq:BBB_Hint_intrinsic} only to single photon states differing in center of mass momentum from ${\bf P}_i$ by the momentum of the photon present; we may therefore simplify the anti-commutator of Eq. \eqref{eq:BBB_Hint_intrinsic} by investigating its action upon this restricted subspace of (zero- or single-photon) states. For example, consider the term
\begin{align}
    &\left\{{\bf P}, \xi_{\bf k} \hat{L}^\dagger_{\bf k} \hat{\pi}^\dagger \hat{a}_{\bf k}\right\} |{\bf P}_i - \hbar {\bf k}, g, {\bf k}\rangle \nonumber \\
    &\hspace{4 em} =\xi_{\bf k}\left(2{\bf P}_i - \hbar {\bf k}\right) |{\bf P}_i, e, 0\rangle \nonumber \\
    &\hspace{4 em} =\xi_{\bf k} \left(2{\bf P}_i - \hbar {\bf k}\right) \hat{L}^\dagger_{\bf k} \hat{\pi}^\dagger \hat{a}_{\bf k} |{\bf P}_i - \hbar {\bf k}, g, {\bf k}\rangle,
\end{align}
where $e$ or $g$ denote the excited or ground states of the particle and $0$ or $\bf k$ denote the zero photon or single photon of mode ${\bf k}$ states. A similar result holds for the second term in the anti-commutator applied to a state $|\mathbf{P}_i, e, 0\rangle$. For a relativistic system the photon momentum is much less than the initial particle momentum, i.e., $\hbar\mathbf{k}\ll\mathbf{P}_i$. Recognizing that ${\bf P}_i=\gamma M {\bf V}_i$ we may thus write for the interaction term,
\begin{align}
    \begin{split}
        \hat{V} = \sum_{\bf k} &\left[ -\left(\frac{1}{\gamma}\vec{d}'_\parallel+\vec{d}'_\perp\right)\cdot \epsilon_{\bf k} + \left(\vec{d}'\times \epsilon^\perp_{\bf k}\right)\cdot \frac{{\bf V}_i}{c} \right] \\
        &\times \left(\xi_{\bf k} \hat{L}^\dagger_{\bf k} \hat{\pi}^\dagger \hat{a}_{\bf k} + \xi^*_{\bf k} \hat{L}_{\bf k} \hat{\pi} \hat{a}^\dagger_{\bf k} \right). \label{eq:BBB_Hint_ac_simp}
    \end{split}
\end{align}
This is the form of the BBB interaction term used within the paper.

\section{\label{app:diagram_teo}Diagrammatic Representation of the Time-Evolution Operator}

This appendix follows closely the standard procedures of the literature; see, for example, the summary work of \cite{Benedict1996} based upon the original papers \cite{Schuurmans1974, Smirnov1973, Sokolov1974}. We work with the state of the system $|\tilde\Psi\left(t\right)\rangle$ in the interaction picture defined as
\begin{equation}
    |\tilde\Psi\left(t\right)\rangle \equiv e^{i \hat{H}_0 (t-t_0) / \hbar} |\Psi\left(t\right)\rangle.
\end{equation}
The state of the system at time $t$ is related to its original state at time $t_0$ via the interaction picture time evolution operator $\widehat{\tilde U}\left(t,t_0\right)$; i.e.,
\begin{equation}
    |\tilde\Psi\left(t\right)\rangle = \widehat{\tilde U}\left(t,t_0\right) |\tilde\Psi\left(t_0\right)\rangle.
\end{equation}
The Schr\"{o}dinger equation for $\tilde U\left(t,t_0\right)$ in the interaction picture reads
\begin{equation}
    i \hbar \frac{\textrm{d} \widehat{\tilde U}}{\textrm{d}t} = \widehat{\tilde V} \widehat{\tilde U} \label{eq:tev_U}
\end{equation}
where
\begin{equation}
    \widehat{\tilde V}(t) = e^{i \hat{H}_0 \left(t-t_0\right)} \hat{V} e^{-i \hat{H}_0 \left(t-t_0\right)}. \label{eq:tev_intpic}
\end{equation}
We require in this work the exact first-order recurrence relation
\begin{align}
    \widehat{\tilde U} \left(t,t_0\right) &= 1 + \left[-\frac{i}{\hbar}\right] \int_{t_0}^t \mathrm{d}t_1 \widehat{\tilde V}\left(t_1\right) \widehat{\tilde U} \left(t_1,t_0\right), \label{eq:tev_o1}
\end{align}
as well as the exact second-order recurrence relation
\begin{align}
    \widehat{\tilde U} \left(t,t_0\right) &= 1 + \left[-\frac{i}{\hbar}\right] \int_{t_0}^t \mathrm{d}t_1 \widehat{\tilde V}\left(t_1\right) \nonumber \\
    &+ \left[-\frac{i}{\hbar}\right]^2 \int_{t_0}^t \mathrm{d}t_2 \int_{t_0}^{t_2} \mathrm{d}t_1 \widehat{\tilde V}\left(t_2\right) \widehat{\tilde V}\left(t_1\right) \widehat{\tilde U} \left(t_1,t_0\right). \label{eq:tev_o2}
\end{align}
We also make use of the infinite series expansion
\begin{align}
    \widehat{\tilde U} \left(t,t_0\right) &= 1 + \left[-\frac{i}{\hbar}\right] \int_{t_0}^t \mathrm{d}t_1 \widehat{\tilde V}\left(t_1\right) \nonumber \\
    &+ \left[-\frac{i}{\hbar}\right]^2 \int_{t_0}^t \mathrm{d}t_2 \int_{t_0}^{t_2} \mathrm{d}t_1 \widehat{\tilde V}\left(t_2\right) \widehat{\tilde V}\left(t_1\right) + \dots \label{eq:tev_series}
\end{align}

A matrix element $\tilde U_{n m} = \langle E_n | \widehat{\tilde U} | E_m\rangle$ coupling two free theory energy eigenstates $|E_n\rangle$ and $|E_m\rangle$ may be interpreted by momentarily leaving the interaction picture. The composition of adjacent occurrences of $\widehat{\tilde V}$ within each integrand translates to the $\left(k,l\right)^\textrm{th}$ non-interaction picture matrix element
\begin{align}
    \widehat{\tilde V}\left(t_{n+1}\right) \widehat{\tilde V}\left(t_{n}\right) &\rightarrow e^{i\omega_k t_{n+1}} V_{k p} e^{-i\omega_p t_{n+1}} e^{i\omega_p t_n} V_{p l} e^{-i\omega_l t_n} \nonumber \\
    &= e^{i\omega_k t_{n+1}} V_{k p} F_p \left(t_{n+1}, t_n\right) V_{p l} e^{-i\omega_l t_n} \label{eq:diag_seq_ex}
\end{align}
where summation over $p$ is implied and where $\hbar \omega_q$ is the energy of a free theory energy eigenstate $|E_q\rangle$. The term
\begin{equation}
    F_p \left(t_{n+1}, t_n\right) = e^{-i\omega_p \left(t_{n+1} - t_n\right)}
\end{equation}
is the free theory propagator of the system in state $|E_p\rangle$ from time $t=t_n$ to time $t=t_{n+1}$.

The sequence $V_{k p} F_p \left(t_{n+1}, t_n\right) V_{p l}$ within expression \eqref{eq:diag_seq_ex} is interpreted from right to left as the physical process wherein the system (i) transitions from state $|E_l\rangle$ to state $|E_p\rangle$ at time $t_n$, (ii) propagates freely in state $|E_p\rangle$ from time $t_n$ to time $t_{n+1}$, and (iii) transitions from state $|E_p\rangle$ to state $|E_k\rangle$ at time $t_{n+1}$. Each term in Eq. \eqref{eq:tev_series} may be interpreted in this manner, where the order of a term in the expansion corresponds to the number of transition events. For each event order sequence, we are directed to integrate over every possible timing of the transition events which satisfies the ordering $t>t_n>t_{n-1}>\dots>t_0$. The summation over transition matrix elements during matrix multiplication of adjacent $V_{kl}$ terms directs us to sum over all possible intermediate transition states.

The particular form of the interaction of interest in this work, given by Eq. \eqref{eq:BBB_Hint_ac_simp}, yields transitions from a state $|e, 0\rangle$ (excited particle with no photon in the radiation field) only to states of the form $|g, {\bf k}\rangle$ (ground state particle with one photon of mode ${\bf k}$ in the radiation field); and transitions from a state of the form $|g, {\bf k}\rangle$ only to the state $|e, 0\rangle$. This interaction therefore greatly restricts the collection of possible transition event sequences; for example, a transition event of the form $|e, {\bf k}\rangle \rightarrow |e, {\bf k}'\rangle$ for ${\bf k} \neq {\bf k}'$ is strictly forbidden \footnote{We refer in such statements to single order processes; in this case, for example, a higher order process involving $|e\rangle \rightarrow |g\rangle \rightarrow |e\rangle$ particle energy level transitions could still yield such a transition}.

\begin{figure}
    \centering
    \includegraphics[width=.75\columnwidth, trim=0cm 0cm 0cm 0cm, clip]{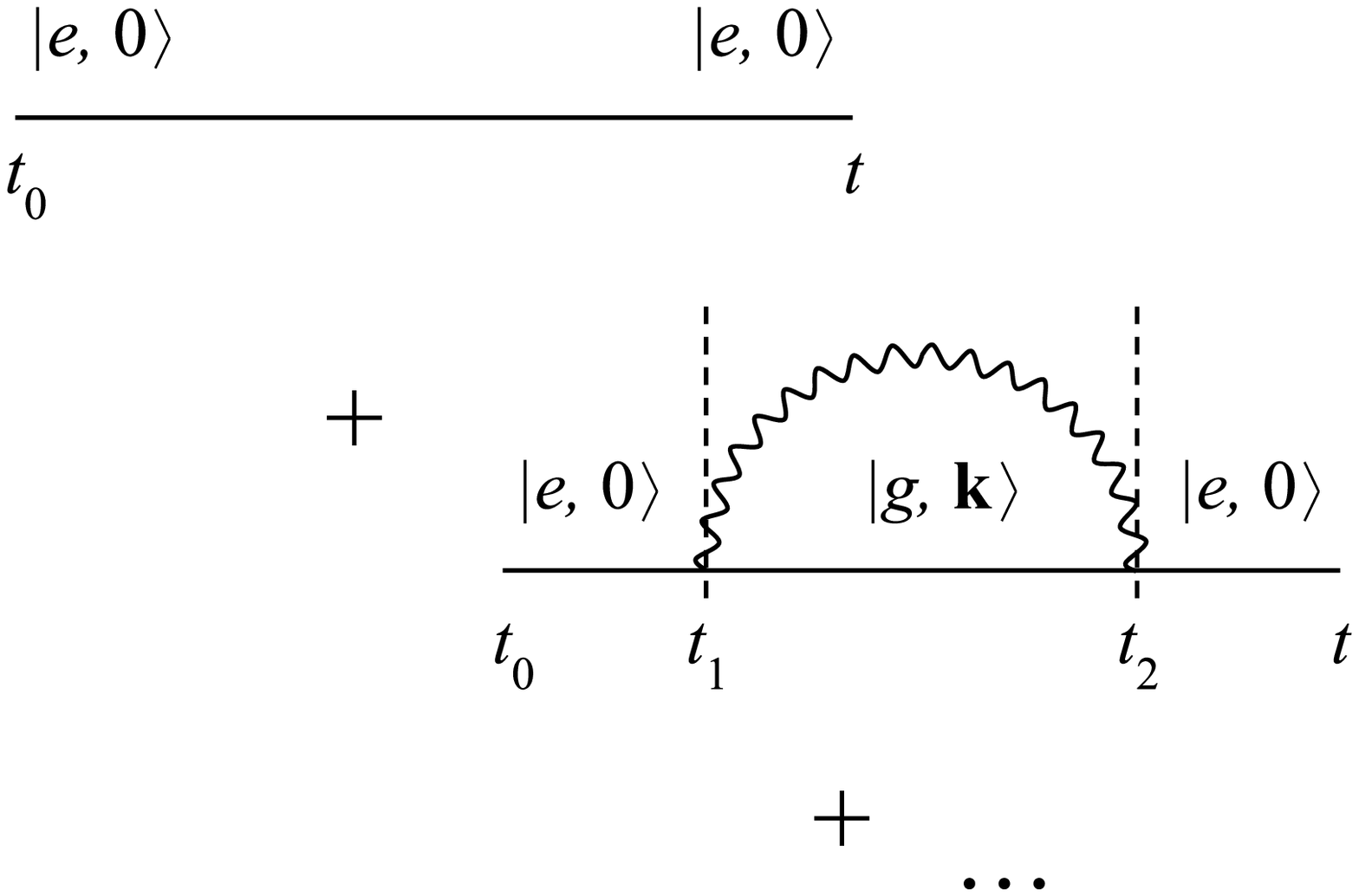}
    \caption{The first two non-vanishing terms in the diagrammatic representation of the time evolution operator matrix element $\tilde U_{e,0;e,0}$. Transition events in the second order term occur at the vertical dashed lines.}
    \label{fig:spont_em_diag}
\end{figure}

Terms in the series expansion of a time evolution operator matrix element are conventionally brought into correspondence with diagrams depicting sequences of transition events. In light of our previous comments, all possible transitions of a single particle system evolving with the interaction defined by Eq. \eqref{eq:BBB_Hint_ac_simp} may be found in Fig. \ref{fig:spont_em_diag}, which shows the first two non-vanishing terms in the diagrammatic representation of the matrix element corresponding to an initially excited particle being found in the excited state after time $t-t_0$.

\begin{widetext}
\section{Calculation of the Relativistic Velocity Coherence Kernel\label{app:conv_viol_term}}

We evaluate in this section the convolution violating term appearing in Eq. \eqref{eq:U11_sr}:
\begin{align}
    &\left[-\frac{i}{\hbar}\right]^2 \int_{t_0}^t \mathrm{d}t_2 \int_{t_0}^{t_2} \mathrm{d}t_1 \sum_{\bf k} \tilde{V}_{1,0;g,{\bf k}}\left(t_2\right) \tilde{V}_{g,\mathbf{k};2,0}\left(t_1\right) \tilde{U}_{2;1}\left(t_1,t_0\right) \nonumber \\
    &\quad =\left[-\frac{i}{\hbar}\right]^2 \int_{t_0}^t \mathrm{d}t_2 \int_{t_0}^{t_2} \mathrm{d}t_1 \sum_{\bf k} e^{i\left(\omega'_0/\gamma - \alpha_{{\bf k}1}\omega_{\bf k}\right) t_2} V_{1,0;g,{\bf k}}\left(t_2\right) e^{-i\left(\omega'_0/\gamma - \alpha_{{\bf k}2}\omega_{\bf k}\right) t_1} V_{g,{\bf k};2}\left(t_1\right) \tilde{U}_{2;1}\left(t_1,t_0\right).
\end{align}
by our prescription of Eq. \eqref{eq:mom_cons}. Let us assume that both dipoles are aligned with the $x$ axis, that both velocities are in the $z$ direction, and that the mode ${\bf k}$ has length $k$ and direction $\Omega$ (with polar and azimuthal parts). Suppressing for a moment the $t_1,t_2$ integrals and $\tilde{U}_{2;1}$, the sum over ${\bf k}$ (including the leading $[-i/\hbar]^2$ multiplier) becomes
\begin{align}
    &-\frac{\Gamma'_0 \gamma c^4}{{\omega'_0}^3} \frac{3}{16\pi^2} \int \mathrm{d} \Omega \left[\cos^2\phi\left(\beta - \cos\theta\right)^2 + \sin^2 \phi \left(1-\beta \cos\theta\right)^2 \right] \int_0^\infty \mathrm{d}k k^3 e^{i\left(\omega'_0/\gamma-\bar{\omega}_{\bf k}\right)\left(t_2-t_1\right)} e^{i k \Delta v \cos\theta \left(t_1+t_2\right)/2} \nonumber \\
    &\quad =-\frac{\Gamma'_0 \gamma c^4}{{\omega'_0}^3} \frac{3}{16\pi} \int_{-1}^1 \mathrm{d}x \left[\left(1+\beta^2\right)\left(1+x^2\right)-4 \beta x \right] \int_0^\infty \mathrm{d}k k^3 e^{i\left(\omega'_0/\gamma-\bar{\omega}_{\bf k}\right)\left(t_2-t_1\right)} e^{i k \Delta v x \left(t_1+t_2\right)/2}
\end{align}
where $\bar{\omega}_{\bf k}=\omega_{\bf k} \alpha_{\bf k}$ with $\alpha_{\bf k}=1-\beta\cos\left(\theta\right)$ for $\theta$ the angle between the mode propagation direction and mean velocity ${\bf v}=\left({\bf v}_1+{\bf v}_2\right)/2$ (or $|{\bf v}| = \beta c$). The velocity difference is $\Delta{\bf v}=\mathbf{v}_2-\mathbf{v}_1$.

We now assume that the integration over $k$ contributes predominantly near where $\omega'_0/\gamma-\bar{\omega}_{\bf k}=0$. We can therefore pull the slowly-varying part $k^3 e^{i k \Delta v x \left(t_1+t_2\right) / 2}$ out of the integral over $k$ and evaluate it at $k'_0/\left(\gamma \alpha_{\bf k}\right)$, make a change of integration variable, and formally extend our integration limits to obtain,
\begin{align}
    &=-\Gamma'_0 \gamma \frac{3}{16\pi} \int_{-1}^1 \mathrm{d}x \frac{\left(1+\beta^2\right)\left(1+x^2\right)-4 \beta x}{\gamma^4 \alpha_{\bf k}^4} e^{i k'_0 \Delta v x \left(t_1+t_2\right)/\left(2\gamma \alpha_{\bf k}\right)} \int_{-\infty}^{\infty} \mathrm{d}\bar{\omega} e^{-i\bar{\omega}\left(t_2-t_1\right)} \nonumber \\
    &=-\frac{\Gamma'_0} {\gamma^3} \frac{3}{8} \int_{-1}^1 \mathrm{d}x \frac{\left(1+\beta^2\right)\left(1+x^2\right)-4 \beta x}{\left(1-\beta x\right)^4} e^{i k'_0 \Delta v x \left(t_1+t_2\right)/\left[2\gamma \left(1-\beta x\right)\right]} \delta\left(t_2-t_1\right)
\end{align}

Plugging this result back into our time integrals above, and recalling that $t_1$ only integrates up to $t_2$ (and thus only picks up half of the support of the Dirac delta distribution), we can immediately execute the integral over $t_1$ to obtain
\begin{equation}
    -\frac{\Gamma'_0} {\gamma^3} \frac{3}{16} \int_{t_0}^t \mathrm{d}t' \int_{-1}^1 \mathrm{d}x \frac{\left(1+\beta^2\right)\left(1+x^2\right)-4 \beta x}{\left(1-\beta x\right)^4} e^{i k'_0 \Delta v x t'/\left[\gamma \left(1-\beta x\right)\right]} \tilde{U}_{2;1}\left(t',t_0\right),
\end{equation}
which is our desired expression.

\end{widetext}

\section{\label{app:photon_em_propagator}Photon Emission Propagator for the BBB Hamiltonian}

We provide in this appendix the single particle propagators corresponding to evolution into non-zero photon states in the BBB Hamiltonian. These propagators are not necessary for the relativistic coherence study of the present work, but would be essential to potential future research into the emission spectrum from a relativistic SR system. We require for this purpose the elements
\begin{align}
    V_{g,{\bf k}; e,0} &= \langle g,{\bf k} | V | e, 0 \rangle \nonumber \\
    &= \left[ -\left(\frac{1}{\gamma}\vec{d}'_\parallel+\vec{d}'_\perp\right)\cdot \epsilon_{\bf k} + \left(\vec{d}'\times\epsilon^\perp_{\bf k}\right)\cdot \frac{{\bf V}_i}{c} \right] \xi^*_{\bf k}.
\end{align}

Let us consider first the case where ${\bf V}_i$ is parallel to $\vec d'$. Using the $\epsilon_{\bf k \theta}$ and $\epsilon_{\bf k \phi}$ polarizations of Section \ref{subsec:self_en_bbb}, only the former yields a non-vanishing interaction matrix element
\begin{equation}
    V^\parallel_{g,{\bf k} \theta; e,0} = \xi^*_{\bf k} d' \sin\theta_{\bf k} / \gamma
\end{equation}
(where $V^\parallel$ denotes that this matrix element is computed for ${\bf V}_i \parallel \vec d'$) or, in the interaction picture,
\begin{align}
    \tilde{V}^\parallel_{g,{\bf k} \theta; e,0} &= e^{i\omega_{g,{\bf k}\theta}t} \left(\xi^*_{\bf k} d' \sin\theta_{\bf k} / \gamma \right) e^{-i\omega_{e,0}t} \nonumber \\
    &= e^{-i\left(\omega'_0/\gamma-\alpha_{\bf k}\omega_{\bf k}\right)t} \xi^*_{\bf k} d' \sin\theta_{\bf k} / \gamma
\end{align}
which has Laplace transform
\begin{equation}
    \tilde{V_\textrm{Q}}^\parallel_{g,{\bf k} \theta; e,0} = \frac{\xi^*_{\bf k} d' \sin\theta_{\bf k} / \gamma}{s+i\left(\omega'_0/\gamma-\alpha_{\bf k}\omega_{\bf k}\right)}.
\end{equation}

The emission matrix element $\tilde U_{g,{\bf k}\theta;e,0}$ of the time evolution operator may be found from the first order exact relation,
\begin{align}
    &\tilde{U}^\parallel_{g,{\bf k}\theta;e,0}\left(t,t_0\right) \nonumber \\
    &\quad = -\frac{i}{\hbar} \int_{t_0}^{t} \mathrm{d}t_1 \sum_m \tilde{V}^\parallel_{g,{\bf k}\theta; m}\left(t_1\right) \tilde U_{m; e, 0}\left(t_1,t_0\right).
\end{align}
The sum over $m$ picks up only the $|e,0\rangle$ state, yielding
\begin{align}
    \tilde{U}^\parallel_{g,{\bf k}\theta;e,0}\left(t,t_0\right) &= -\frac{i}{\hbar} \int_{t_0}^{t} \mathrm{d}t_1 \tilde{V}^\parallel_{g,{\bf k}\theta;e,0}\left(t_1\right) \tilde U_{e,0;e,0}\left(t_1,t_0\right)\label{eq:tildeu_gke}
\end{align}
which becomes in the Laplace domain
\begin{align}
    \tilde{\mathcal{U}}^\parallel_{g,{\bf k}\theta;e,0} &= -\frac{i}{\hbar} \left[\frac{\xi^*_{\bf k} d' \sin\theta_{\bf k}/\gamma}{s+i\left(\omega'_0/\gamma-\alpha_{\bf k}\omega_{\bf k}\right)}\right] \left[\frac{1}{s+\Gamma'_0/\left(2\gamma\right)}\right] \nonumber \\
    &= -\frac{i}{\hbar} \left[\frac{\xi^*_{\bf k} d' \sin\theta_{\bf k}/\gamma}{\Gamma'_0/\left(2\gamma\right)+i\left(\alpha_{\bf k}\omega_{\bf k}-\omega'_0/\gamma\right)}\right] \nonumber \\
    &\quad \times \left[ \frac{1}{s+i\left(\omega'_0/\gamma-\alpha_{\bf k} \omega_{\bf k}\right)} - \frac{1}{s+\Gamma'_0/\left(2\gamma\right)} \right],
\end{align}
such that in the time domain,
\begin{align}
    \tilde{U}^\parallel_{g,\mathbf{k}\theta;e,0}\left(t,t_0\right) &= -\frac{i}{\hbar} \left[\frac{\xi^*_{\bf k} d' \sin\theta_{\bf k}/\gamma}{\Gamma'_0/\left(2\gamma\right)+i\left(\alpha_{\bf k}\omega_{\bf k}-\omega'_0/\gamma\right)}\right] \nonumber \\
    &\qquad \times \left[e^{i\left(\alpha_{\bf k} \omega_{\bf k}-\omega'_0/\gamma\right) \Delta t} - e^{-\Gamma'_0 \Delta t /\left(2\gamma\right)} \right]. \label{eq:Ugk_prop_par}
\end{align}
In the limit as $t\rightarrow\infty$ the probability of a photon having been emitted of mode ${\bf k}\theta$ is
\begin{equation}
    \mathcal{P}^\parallel_{{\bf k}\theta}\left(t=\infty\right) = \frac{|\xi_{\bf k}|^2}{\hbar^2} \left[\frac{d'^2 \sin^2\theta_{\bf k}/\gamma^2}{\left(\omega'_0/\gamma-\alpha_{\bf k}\omega_{\bf k}\right)^2 +\Gamma'^2_0/\left(2\gamma\right)^2}\right]
\end{equation}
which is a Lorentzian centered at the expected relativistic Doppler shifted frequency $\omega = \omega'_0 / \left[\gamma\left(1-\cos\theta V/c\right)\right]$. Although this result constitutes another validation of our methods and of the BBB Hamiltonian proposed in \cite{Boussiakou2002}, it is the propagator of Eq. \eqref{eq:Ugk_prop_par} that is of use to future researchers. Note that the arbitrary fiducial quantisation volume present in $|\xi_{\bf k}|^2$ may be removed by changing over to a mode probability density in $\left\{\Omega,\omega\right\}$-space and working with the physically significant intensity, proportional to the probability density per solid angle.

The above was computed for the case where ${\bf V}_i$ was parallel to $\vec d'$; if ${\bf V}_i$ is now perpendicular to $\vec d'$, we obtain the propagators
\begin{align}
    \tilde{U}^\perp_{g,\mathbf{k}\theta;e,0}\left(t,t_0\right) &= -\frac{i}{\hbar} \left[\frac{\xi^*_{\bf k} d' \left(\sin\theta_{\bf k}-\cos\phi_{\bf k} V_i/c\right) }{\Gamma'_0/\left(2\gamma\right)+i\left(\alpha_{\bf k}\omega_{\bf k}-\omega'_0/\gamma\right)}\right] \nonumber \\
    &\qquad \times \left[e^{i\left(\alpha_{\bf k} \omega_{\bf k}-\omega'_0/\gamma\right) \Delta t} - e^{-\Gamma'_0 \Delta t /\left(2\gamma\right)} \right] \label{eq:Ugk_prop_perp_theta} \\
    \tilde{U}^\perp_{g,\mathbf{k}\phi;e,0}\left(t,t_0\right) &= -\frac{i}{\hbar} \left[\frac{\xi^*_{\bf k} d' \cos\theta_{\bf k}\sin\phi_{\bf k}}{\Gamma'_0/\left(2\gamma\right)+i\left(\alpha_{\bf k}\omega_{\bf k}-\omega'_0/\gamma\right)}\right] \nonumber \\
    &\qquad \times \left[e^{i\left(\alpha_{\bf k} \omega_{\bf k}-\omega'_0/\gamma\right) \Delta t} - e^{-\Gamma'_0 \Delta t /\left(2\gamma\right)} \right] \label{eq:Ugk_prop_perp_phi}
\end{align}
where $\phi_{\bf k}$ is the angle between ${\bf V}_i$ and the projection of $\bf k$ onto the plane perpendicular to $\vec d'$.

% \nocite{*}

\bibliography{SR_Relativistic_CWyenberg_APS}% Produces the bibliography via BibTeX.

\end{document}